\documentclass[aoas]{imsart}

\RequirePackage{amsthm,amsmath,amsfonts,amssymb}
\RequirePackage[authoryear]{natbib}
\usepackage{algorithm}
\usepackage{algpseudocode}
\usepackage{graphicx}
\usepackage[usenames,dvipsnames]{xcolor}
\usepackage[colorlinks=true, allcolors=blue]{hyperref}
\usepackage{multirow}
\usepackage{booktabs}
\usepackage{graphicx} 
\usepackage[english]{babel}
\usepackage{natbib}
\usepackage[capitalize,noabbrev]{cleveref}

\usepackage{subcaption}
\usepackage{bm}

\usepackage{array}
\usepackage{ragged2e}
\newcolumntype{Y}{>{\centering\arraybackslash}X}
\newcolumntype{P}{>{\raggedleft\arraybackslash}X}
\newcolumntype{C}[1]{ >{\centering\arraybackslash} m{#1} }
\newcolumntype{R}[1]{ >{\raggedright\arraybackslash} m{#1} }

\newcommand{\bW}{\mathbf{W}}
\newcommand{\bU}{\mathbf{U}}
\newcommand{\bw}{\mathbf{w}}
\newcommand{\bu}{\mathbf{u}}

\newcommand{\bY}{\mathbf{Y}}

\newcommand{\bM}{\mathbf{M}}
\newcommand{\bZ}{\mathbf{Z}}

\newcommand{\bC}{\mathbf{C}}

\newcommand{\cL}{\mathcal{L}}
\newcommand{\cN}{\mathcal{N}}
\newcommand{\cO}{\mathcal{O}}
\newcommand{\EE}{\mathbb{E}}
\newcommand{\PP}{\mathbb{P}}
\newcommand{\RR}{\mathbb{R}}

\newcommand{\op}{o_{\PP}}
\newcommand{\Op}{\cO_{\PP}}

\newcommand{\bbeta}{\bm{\beta}}
\newcommand{\diag}{\mathrm{diag}}

\newtheorem{theorem}{Theorem}[section]

\newtheorem{remark}[theorem]{Remark}
\newtheorem{assumption}[theorem]{Assumption}

\usepackage[toc,page]{appendix}

\begin{document}

\begin{frontmatter}
\title{Augmented doubly robust\\post-imputation inference for proteomic data}
\runtitle{Post-Imputation Inference}
\runauthor{H.MOON, J.DU, J.LEI and K.ROEDER}

\begin{aug}
\author{\fnms{Haeun}~\snm{Moon}\ead[label=e1]{haeunmoon@snu.ac.kr}},
\author{\fnms{Jin-Hong}~\snm{Du}\ead[label=e2]{jinhongd@andrew.cmu.edu}},
\author{\fnms{Jing}~\snm{Lei}\ead[label=e3]{jinglei@andrew.cmu.edu}},
\and
\author{\fnms{Kathryn}~\snm{Roeder}\ead[label=e4]{roeder@andrew.cmu.edu}}

\address{Department of Statistics, Seoul National University \printead[presep={,\ }]{e1}}

\address{Department of Statistics and Data Science,
Carnegie Mellon University\printead[presep={,\ }]{e2,e3,e4}}

\end{aug}

\begin{abstract}
Quantitative measurements produced by mass spectrometry proteomics experiments offer a direct way to explore the role of proteins in molecular mechanisms. However, analysis of such data is challenging due to the large proportion of missing values. A common strategy to address this issue is to utilize an imputed dataset, which often introduces systematic bias into downstream analyses if the imputation errors are ignored. In this paper, we propose a statistical framework inspired by doubly robust estimators that offers valid and efficient inference for proteomic data. Our framework combines powerful machine learning tools, such as variational autoencoders, to augment the imputation quality with high-dimensional peptide data, and a parametric model to estimate the propensity score for debiasing imputed outcomes. Our estimator is compatible with the double machine learning framework and has provable properties. Simulation studies verify its empirical superiority over other existing procedures. In application to both single-cell proteomic data and bulk-cell Alzheimer's Disease data our method utilizes the imputed data to gain additional, meaningful discoveries and yet maintains good control of false positives. 
\end{abstract}



\begin{keyword}
\kwd{proteomic data}
\kwd{post-imputation inference}
\kwd{double robustness}
\kwd{variational autoencoder}
\end{keyword}

\end{frontmatter}

\section{Introduction}
Recently single-cell RNA sequencing technology has fueled a revolution in our ability to study biological processes. However, mRNA transcript abundances are only a weakly correlated precursor to protein abundances \citep{Vogel:2012th, Liu:2016vj,Tasaki:2022td}. And it is the protein that carries out the more fundamental roles of molecular mechanisms in cellular processes.
Developments in mass spectrometry proteomic technology have greatly enhanced the quantitative analysis of proteins related to human health and disease. Nevertheless, such analyses often encounter challenges due to a high rate of missingness, especially for single-cell data, resulting from various technological factors \citep{vanderaa2023revisiting}.  While missingness significantly impacts the validity and efficiency of downstream tasks, the optimal method for handling missing data in proteomics remains a subject of active debate, and is an area in need of novel methodological advancements \citep{shen2022comparative}.

To assess protein abundance, the measured units are peptides, which are subunits of a protein.  Peptides that are present in the sample matrix, but not assigned an abundance value for the observations in a batch are considered missing. Missingness can be attributed to a variety of technical factors that lead to a failure to measure abundance across all observed spectra \citep{Webb-Robertson:2015, Bramer:2021,vanderaa2023revisiting}. Missing patterns have been reported to be close to random \citep{Brenes:2019}, meaning that while the propensity depends weakly on measured covariates, it tends not to depend on the true value of the abundance. Peptides with very low abundance are more likely to be missing; however, by design, mass spectrometry is calibrated to measure common peptides and hence no measurements are made on the majority of low abundance peptides.  
Thus while some missing values are missing not at random (MNAR), the vast majority follow the missing at random (MAR) assumption.

Currently, a major focus of discussion in the field is on the choice of imputation method \citep{vanderaa2023revisiting, wei2018missing}, which is used to infer peptide abundance. It is common practice that imputed values are directly plugged into the original dataset, followed by downstream analyses as if the imputed values were the original observed ones (``Plugin method''). With this method, the assumption is that the imputed data accurately represents the original data. Therefore, the precision of the imputed result is crucial for a valid downstream analysis. A substantial ongoing research effort is to search and experiment with numerous imputation methods to determine the optimal one, including sample matching methods \citep{stuart2019integrative}, matrix factorization methods \citep{hastie2015matrix}, deep learning methods \citep{yoon2018gain,qiu2020genomic, Du:2022} and more \citep{wang2016imputing, chen2017mixed}. See \cite{harris2023evaluating, valikangas2018comprehensive, liu2021proper} for a comprehensive review. 

Most of the aforementioned methods rely on a high-dimensionality and robust intercorrelation structure of the measured peptides. Such characteristics of proteomic data provide a solid foundation for various imputation algorithms; however, this approach may not be ideal when the downstream analysis plan is based on the Plugin method. There are two reasons for this. First, the aim of retrieving the original outcomes via imputation is not optimal in some downstream analyses. Consider a linear model in which we regress each peptide abundance in some low-dimensional covariates. In this context, the optimal choice for imputation is the conditional mean abundance based on these covariates. When the Plugin method is combined with high-dimensional imputation models, we are attempting to get closer to the original outcome, rather than the conditional mean, which may introduce additional variance into estimated regression coefficients. Second, when the full high-dimensional dataset is used for imputation, a systematic bias can be introduced into the imputed data, causing false discovery due to confounding.  
A recent paper by \citet{agarwal2020data} investigates this issue using transcriptomic datasets.  They show that if the dataset contains a number of differentially expressed genes, a naive application of the Plugin method results in notably inflated False Discovery Rates (FDR). This inflation does not occur when none of the genes are differentially expressed, which indicates that the source of the FDR inflation is the cross-use of high-dimensional data for imputation. 
More discussions on this can be found in \cite{andrews2018false, ly2022effect}.
Similar post-imputation inference issues remain for proteomic studies.

One approach that can circumvent these issues is to use only complete data for analysis and simply ignore missingness (``Complete method''). This provides a simple and valid way to prevent problems from imputation under certain missingness assumptions. However, it discards any indirect information on missing outcomes and is especially vulnerable to low power with small sample sizes. Multiple imputation \citep{rubin2004multiple} is another possible approach, which provides a general framework for obtaining valid statistical inferences while incorporating the imputation uncertainty. This technique avoids denoising and involves generating multiple complete datasets by filling in missing data with several plausible imputations.  The resulting test statistic incorporates variances both within and between datasets to compute the total variance.  There have been some attempts to apply this framework to proteomic data \citep{yin2016multiple, gianetto2020peptide}.
Some noticeable challenges involved in using this approach include its empirical conservativeness \citep{Chion:2022}, computational burden \citep{brini2023missing}, and the lack of a straightforward expression for test statistics \citep{meng1994multiple}.

In this paper, we propose an alternative framework motivated by doubly robust estimation \citep{scharfstein1999adjusting}, a widely used procedure to estimate mean outcomes. Our purpose is to establish a valid and efficient inference framework that is well-harmonized with high-dimensional imputation models. Estimating mean outcomes is a significant area of research, especially when certain outcomes are not observable and a propensity score (probability of observation) depends on measured covariates. Then observed outcomes do not accurately represent the entire population due to the covariate mismatch. Therefore, instead of simply averaging the observed outcomes, one first constructs an outcome model by regressing the outcomes on covariates related to the propensity score and averaging the fitted values over the entire population. A doubly robust estimator incorporates an additional term to correct for the first-order bias of the fitted outcomes. While two nuisance estimators -- an outcome estimator and a propensity score estimator -- are employed, this approach enjoys a ``double robustness'' property, which means that the statistic remains consistent as long as at least one of the nuisance estimators is consistent \citep{robins1995semiparametric}. Several recent papers extend this strategy to estimation problems beyond the mean outcome \citep{kennedy2023towards, fisher2023three, diaz2018targeted, qiu2023efficient}. In particular, \cite{kennedy2023towards} uses each summand of the doubly robust estimator as a pseudo-outcome to measure a conditional average treatment effect in a nonparametric regression setting.

\begin{figure}[h]
  \centering
    \includegraphics[width=0.5\linewidth]{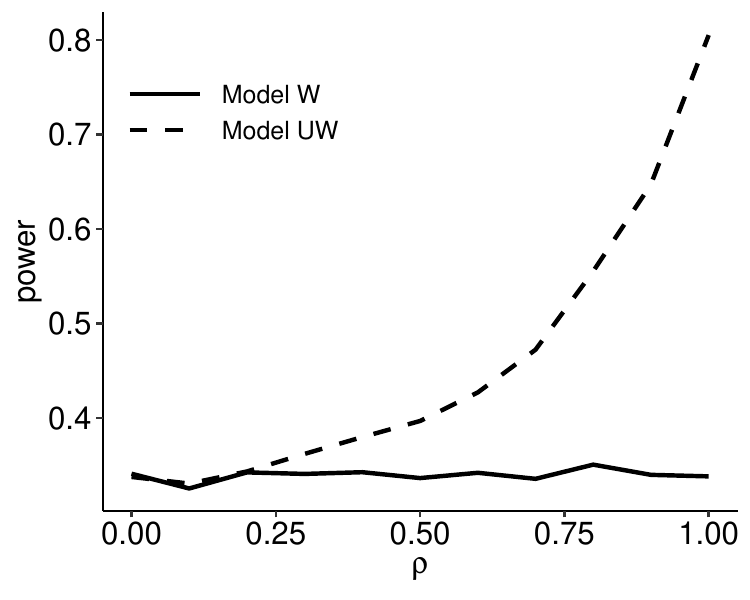}
  \caption{Statistical power of rejecting $\bbeta=0$ at different levels of correlation $\rho\in\{0.1,0.2,\cdots,1\}$ between an outcome $Y_i$ and an auxilary variable $U_i$.  Further implementation details are provided in \Cref{subsec:augment}. 
  }
  \label{fig:toy}
\end{figure}

Adopting this strategy to a linear regression setting, we utilize the summands of the aforementioned doubly robust estimator as pseudo-outcomes and transfer its favorable properties to regression coefficients. Moreover, the availability of high-dimensional proteomic data offers us the opportunity to augment our estimator by using this additional information. We show that the asymptotic variance of the estimated coefficients is further reduced by augmenting the imputation model.  Our strategy is to use the entire proteomic data as an auxiliary variable and use their intercorrelated structure for imputation. To illustrate the usefulness of this approach, we provide a simple experiment.  Assume that there exists an auxiliary variable that is correlated with the outcome of interest. Then the outcome model with the auxiliary variable (Model UW) provides better statistical power compared to a model without it (Model W) in a downstream task, and the gap increases as the auxiliary variable becomes more informative for the outcome variable (Figure \ref{fig:toy}). Further details of implementation and interpretation are provided in \Cref{subsec:augment}.

In our framework, the propensity score is estimated through a conventional logit model to enjoy a fast rate of parametric convergence, while the outcome model is estimated through a flexible machine-learning method that can handle high-dimensional variables and their complex relationships. Our framework not only calls for, but also deliberately invites powerful modern methods because it includes a built-in mechanism to push the estimator towards achieving $\sqrt{n}$-consistency, even when the employed imputation method fails to achieve a sufficiently fast rate. In our simulations and data study, we use a variant of VAE models called VAEIT \citep{Du:2022} to fit the outcomes; see  B for more details. The VAEIT model utilizes both low-dimensional covariates and high-dimensional proteomic data, and offers enough flexibility to handle missing data as well as non-linear dependency.

\paragraph*{Other related works} 
In high-dimensional nuisance parameter estimation, \cite{jiang2022new} and \cite{yadlowsky2022explaining} derived consistency results for estimated conditional treatment effect with sparsity or distributional assumptions. Double machine learning, proposed by \cite{chernozhukov2018double}, provides a framework for building an efficient estimator of low-dimensional parameters, with nuisance functions estimated using a high-dimensional black-box model. More papers based on semiparametric nuisance estimation are summarized in \cite{davidian2022methods}.  Most of the aforementioned references use the same set of high-dimensional variables for both nuisance functions.  Other lines of investigation, including \cite{berrevoets2023impute}, \cite{zhao2022adjust}, and \cite{little2012prevention}, explore an estimator for average treatment effect when some data are missing. They measure the effect size by adjusting the covariate distributions of treatment and control groups separately and computing the outcome difference. In a matrix completion problem, \cite{chen2019inference} suggests a method for denoising a matrix with corrupted and missing entries based on low-rank decomposition. Their method debiases the initial rank-constrained estimator and provides confidence intervals with distributional guarantees. \cite{shao2023distribution}, \cite{gui2023conformalized} suggests entrywise predictive inferences using a conformal prediction framework.
This line of work aims to provide uncertainty quantification for each imputed entry and doesn't consider the associations between the outcomes and the treatment, which is substantially different from our work.

The rest of our paper is organized as follows. In Section \ref{sec:method}, we formally introduce the doubly robust estimator, and our procedure for estimating a regression coefficient drawn from doubly robust pseudo-outcomes. We then motivate the use of augmented imputation, define the augmented doubly robust estimator, and establish its asymptotic properties.  In Section \ref{sec:multiple}, we describe a multiple testing procedure as an example of downstream applications of the proposed estimators and demonstrate their favorable finite sample performance compared to benchmark methods. Next, we apply the proposed method to analyze a real proteomic dataset. In Section \ref{sec:sc}, we analyze a single-cell peptide dataset with cell-specific covariates, identifying peptides whose abundance is related to the cell size.
In Section \ref{sec:AD}, we apply the proposed method to a bulk-cell dataset annotated with a range of Alzheimer’s Disease symptoms. Section \ref{sec:discussion} summarizes the paper and discusses possible issues in the application of the proposed method. The results presented in Section \ref{sec:sc} and \ref{sec:AD} can be reproduced using the code provided at \url{https://github.com/HaeunM/peptide-imputation-inference}.

\section{Method} \label{sec:method}
\subsection{Background} \label{subsec:background}

Suppose $n$ identically and independently distributed samples $(\bW_1, Y_1)$, $\ldots,(\bW_n, Y_n)$ $\in \RR^q\times\RR$ are drawn from a linear model:
\begin{align} 
    Y_i= \bW_i^T\bbeta+ \epsilon_i  , \label{true}
\end{align}
where $\bbeta \in \mathbb{R}^{q}$ is the coefficient vector and $\epsilon_i  \in \mathbb{R}$ is a zero-mean noise.
We consider the missing data problem when some of the outcomes $Y_i$'s are not observable.
Specifically, we denote the observability of $Y_i$ by a binary random variable $C_i\in\{0,1\}$, such that one can only observe $(\bW_i,C_i,C_iY_i)$ for $i=1,\ldots,n$.
Under the missing data setting, we are interested in testing the hypothesis:
\[H_0: \bbeta=0 \text{ \quad versus \quad} H_1: \bbeta \neq 0.\] 

 If every outcome is observable ($C_i=1$ for all $i\in \{1,\cdots,n\}$), the ordinary least square regression (OLS) is arguably the most common method for estimating $\bbeta$: \begin{equation}
     \hat{\bbeta}_{OLS}=\text{arg min}_{\bbeta}\sum_{i=1}^{n}(Y_i-\bW^T_i\bbeta)^2=\left(\sum_{i=1}^n \bW_i\bW^T_i\right)^{-1} \left(\sum_{i=1}^n \bW_i Y_i\right)\,. \label{eq:betaols}
 \end{equation} A test statistic can be obtained based on its asymptotic distribution
 $$\sqrt{n}(\hat{\bbeta}_{OLS}-\bbeta) \xrightarrow{D} \cN(0, \EE[\bW_i\bW^T_i]^{-1}\EE[\epsilon_i^2\bW_i\bW_i^T]\EE[\bW_i\bW^T_i]^{-1}),$$ where the asymptotic covariance can be approximated by a plugin estimator $$\left(\frac{1}{n}\sum_{i=1}^n \bW_i\bW_i^T\right)^{-1}\left(\frac{1}{n}\sum_{i=1}^{n}({Y}_i-\bW_i\hat{\bbeta}_{OLS})^2\bW_i\bW_i^T\right)\left(\frac{1}{n}\sum_{i=1}^n \bW_i\bW_i^T\right)^{-1}.$$ This is one of the most well-known inference frameworks in statistics. 
 When some outcomes are not observed, the least squares estimate is not applicable. 
 
 If the rate of missingness is only related to a measured covariate ($C_i \perp Y_i |\bW_i$), a simple strategy of excluding missing samples provides valid inferential results \citep{little1992regression}; but it comes at the expense of a reduced sample size. Therefore, we consider the pseudo-outcome approach, which can offer better statistical efficiency. In an ideal scenario, when the conditional mean $\mathbb{E}[Y_i\mid \mathbf{W}_i]$ is available, replacing the outcome data with this value will provide valid and efficient inference. This approach has been explored in causal inference studies. While the typical average treatment effect estimates $\mathbb{E}[Y_i]$, the conditional average treatment effect seeks an individualized conditional outcome $\mathbb{E}[Y_i\mid \mathbf{W}_i]$, especially when $Y_i$ is not observable for counterfactual cases. Several recent papers address this issue by utilizing pseudo-outcomes, which have the same conditional means as the original outcomes, and fitting a regression against them as if they were observed data \citep{kennedy2023towards, fisher2023three, semenova2021debiased, diaz2018targeted}.
This approach allows for achieving desirable properties such as robustness and efficiency through a selection of appropriate pseudo-outcomes.

Inspired by these studies, we further extend the pseudo-outcome framework in linear regression. We especially focus on the doubly robust estimator suggested by \cite{scharfstein1999adjusting}, which is extensively used to estimate the mean outcome $\mathbb{E}[Y_i]$. This estimator is defined upon the construction of two nuisance functions;  
\begin{align*}
    \mu(\bw) &= \mathbb{E}[Y_i\mid\mathbf{W}_i=\bw] \tag{Outcome model}\\
    \delta(\bw) &= \mathbb{P}(C_i=1\mid\mathbf{W}_i=\bw), \tag{Propensity model}
\end{align*} and is formulated as $\frac{1}{n}\sum_{i=1}^ng(Y_i, C_i;\hat{\mu}, \hat{\delta})$, where \[g(Y_i, C_i;\mu, \delta)=\mu(\mathbf W_i)+\frac{C_i}{\delta(\mathbf W_i)}(Y_i-\mu(\mathbf W_i)),\] with estimated outcome and propensity models, $\hat\mu$ and $\hat\delta$.
In general, these models can be obtained through parametric or nonparametric regression methods.
Appealing properties of the estimator arise from its second-order nuisance estimation error, that is,  
\begin{equation}\label{eq:bias_2nd}
    \EE[g(Y_i, C_i;\hat{\mu},\hat{\delta})-g(Y_i,C_i;\mu,\delta)\mid \hat\mu,\hat\delta]=(\mu(\mathbf W_i)-\hat{\mu}(\mathbf W_i))\left(1-\frac{\delta(\mathbf W_i)}{\hat{\delta}(\mathbf W_i)}\right).
\end{equation} Then, under some weak assumptions on convergence rates of $\hat\mu$ and $\hat\delta$, the bias of the estimator from the nuisance estimation error becomes negligible \citep{kennedy2023towards}. Moreover, consistency of the estimator is achieved if either $\mu$ or $\delta$ is consistently estimated, which is referred to as the doubly robust property.

In the regression setting, pseudo-outcomes can be introduced as follows:
\[\hat{Y}_i^W = \hat{\mu}_i+\frac{C_i}{\hat{\delta}_i}(Y_i-\hat{\mu}_i). \]
Here, and in the rest of this paper, we write $\mu_i=\mu(\mathbf W_i)$ and $\delta_i=\delta(\mathbf W_i)$ and similarly for the estimated versions. 
Regressing $(\hat{Y}^W_1,...,\hat{Y}^W_n)$ on $(\bW_1,...,\bW_n)$ yields a least squares estimator given by:
$$\hat{\bbeta}_W=\left(\sum_{i=1}^n \bW_i\bW^T_i\right)^{-1} \left(\sum_{i=1}^n \bW_i \hat{Y}_i^W\right).$$ 
As we will show in \Cref{subsec:augment}, the estimator $\hat{\bbeta}_W$ also has the doubly robust property and can lead to more efficient inference. 

\subsection{An augmented doubly robust estimator $\hat{\bbeta}_{UW}$} \label{subsec:augment}
In peptide abundance analysis, there are often a large collection of peptides measured and analyzed together. For each peptide, one can predict its value using not only the low-dimensional covariate $\bW$ but also the other peptides, which can be regarded as a high-dimensional covariate $\bU$.
Our strategy is to recover $Y$ as accurately as possible through an augmented outcome model that incorporates both $\bW$ and $\bU$ as predictors for the response $Y$.  If the augmented outcome model will result in a significant reduction in the variance of the regression residual $Y-\EE[Y|\bW,\bU]$, then we may expect to have a smaller asymptotic variance for the estimated regression coefficient using the augmented pseudo-outcome.

Formally, our proposed estimator is defined as 
\begin{align}
    \hat{\bbeta}_{UW}=\left(\sum_{i=1}^n \bW_i\bW^T_i\right)^{-1} \sum_{i=1}^n \bW_i \left(\hat{\nu_i}+\frac{C_i}{\hat{\delta}_i}(Y_i-\hat{\nu}_i)\right) \label{eq:hbeta-UW}
\end{align}
for nuisance estimators $\hat{\nu}(\bw,\bu)=\hat{\mathbb{E}}[Y_i| \bW_i=\bw, \bU_i=\bu]$ and $\hat{\delta}(\bw)=\hat{\mathbb{E}}[C_i= 1| \bW_i=\bw].$ 

Before providing a rigorous analysis, we provide a simple example to illustrate the variance reduction effect of augmentation. Consider a linear regression model $Y_i=\beta W_i + \epsilon_i$ for $\beta,W_i  \in \mathbb{R}$. An auxilary variable $U_i \in \mathbb{R}$ is defined as $U_i=\beta W_i+{\epsilon}_{u_i}$, where $\mathrm{Cor}(\epsilon_{u_i}, \epsilon_i)=\rho$. Since $U_i$ partly explains the residual term $\epsilon_i$, the outcome $\nu_i=\EE[Y_i \mid W_i,U_i]$ provides a higher resolution estimate of $Y_i$ than $\mu_i=\EE[Y_i\mid W_i]$. We compare two pseudo-outcomes ${\hat Y}^{W}_i =\hat{\mu}_i+\frac{C_i}{\hat{\delta}_i}(Y_i-\hat{\mu}_i)$ (Model W) and ${\hat Y}^{UW}_i =\hat{\nu}_i+\frac{C_i}{\hat{\delta}_i}(Y_i-\hat{\nu}_i)$ (Model UW) in their downstream performance. Specifically, we perform a linear regression against each pseudo-outcome on $\bW_i$ and their statistical powers in rejecting $\bbeta=0$ are compared. The outcome $Y_i$ has random missingness with a known observation probability $\delta_i=0.7$, the true coefficient is $\beta=0.2$, and the sample size is $n=200$. The results are averaged over 5000 repetitions.  The result shows that Model UW outperforms Model W, and it provides increasing power as the auxiliary variable becomes more informative for the outcome ($\rho \to 1$, Fig. \ref{fig:toy}). In real applications, it is less probable that a single protein exhibits such a substantial correlation with an outcome. Instead, high-dimensional proteomic data may collectively contribute to recovering the outcome.  

Next, we derive asymptotic properties of the proposed estimator $\hat{\bbeta}_{UW}$ rigorously. Here, we prove that $\hat{\bbeta}_{UW}$ possesses a doubly robust property (Theorem \ref{thm:DR}) and asymptotic normality (Theorem \ref{thm:asymp_norm}), and its asymptotic variance is smaller than that of $\hat{\bbeta}_{W}$ (Theorem \ref{thm:high_better}).

\vspace{2mm}
\textit{Notation.} We denote the $L_2$ norm of a vector, or a random variable, or a function of a random variable as $\|\cdot\|_2$. For example, for a random vector $\bW$ and its function $\nu=\nu(\bW)$, $\|\nu\|_2$ is defined as $(\int \|\nu(\bW)\|_2^2 \mathrm{d} P_W )^{1/2}$.  $L$-infinity norm of a vector, or a random variable, or a function of a random variable is denoted as  $\|\cdot\|_{\infty}$. For matrices $M_A$ and $M_B$, we write as $M_A \preccurlyeq M_B$ if $(M_B-M_A)$ is positive semidefinite.

\begin{assumption} \label{assumption}
        \begin{itemize}   
                \item [(a)] Missing at random : ${Y_i} \perp C_i \mid (\bW_i, \bU_i)$ 
	\item [(b)] The propensity score: $\delta(\bW_i)=\PP(C_i=1\mid  \bW_i)=\PP(C_i=1\mid  \bW_i, \bU_i) \in (0,1]$ is bounded away from 0 by some constant with probability 1.
	\item [(c)] Noise : $\EE[\epsilon_i\mid  \bW_i]=0$, {$\EE[Y_i-\nu_i\mid  \bW_i, \bU_i]=0$}, $\|\epsilon_i\|_2$ and $\|Y_i-\nu_i\|_{\infty}$ are bounded. 
	\item [(d)] Covariate :  $\|\bW_{i}\|_\infty $ is bounded, $\EE[\bW_i \bW_i^T]$ is a full-rank matrix.
        \end{itemize}
\end{assumption}
The second equality of Assumption \ref{assumption}(b) requires conditional independence between $C$ and $\mathbf U$ given $\mathbf W$. This is the key assumption that allows us to use an augmented outcome model to improve efficiency.

Under the above assumptions and some additional mild assumptions on nuisance estimations, the doubly robust property follows, as shown in the following theorem.

\begin{theorem}[Double robustness]\label{thm:DR} 
Assume \Cref{assumption} (a)-(d). If one of the nuisance parameters is consistent, i.e., $\| \frac{\delta_i}{\hat{\delta_i}} -1\|_2=\op(1)$ or $\|\hat{\nu_i}-\nu_i\|_2=\op(1)$,
then the estimator $\hat{\bbeta}_{UW}$ defined in \eqref{eq:hbeta-UW} is consistent, i.e., $\hat{\bbeta}_{UW} \xrightarrow{P} \bbeta$. 
\end{theorem}

\Cref{thm:DR} guarantees the consistency of the proposed estimator.
If further, the product of the nuisance estimation errors is small, we can derive the asymptotic distribution of $\hat{\bbeta}$.

\begin{theorem}[Asymptotic normality]\label{thm:asymp_norm} 
    Under the same conditions in \Cref{thm:DR}, further assume that both of the nuisance parameters are consistent, and $\|(1-{\delta_i}/{\hat{\delta_i}})(\hat{\nu_i}-\nu_i)\|_2=\op(n^{-1/2})$. 
    Then the estimator $\hat{\bbeta}_{UW}$ defined in \eqref{eq:hbeta-UW} is asymptotically normal:
     \begin{equation*}
     \sqrt{n}(\hat{\bbeta}_{UW}-\bbeta) \xrightarrow{D} \cN(0, \Sigma_{UW})
     \end{equation*}
     where $\Sigma_{UW}=\EE[\bW_i\bW^T_i]^{-1}\EE[(\epsilon_i^2+(\frac{1}{\delta_i}-1)(Y_i-\nu_i)^2)\bW_i\bW_i^T]\EE[\bW_i\bW^T_i]^{-1}$. The asymptotic covariance $\Sigma_{UW}$ can be consistently estimated by a plug-in estimator 
    \begin{equation}
        \hat{\Sigma}_{UW}=\left(\frac{1}{n}\sum_{i=1}^n \bW_i\bW_i^T\right)^{-1}\left(\frac{1}{n}\sum_{i=1}^{n}(\hat{Y}_i^{UW}-\bW_i^T\hat{\bbeta}_{UW})^2\bW_i\bW_i^T\right)\left(\frac{1}{n}\sum_{i=1}^n \bW_i\bW_i^T\right)^{-1}
        \label{eq:hetero}
    \end{equation}
\end{theorem}

The asymptotic variance $\Sigma_{UW}$ in Theorem \ref{thm:asymp_norm} is identical to the variance obtained with oracle nuisance functions. That is, when both nuisance estimates are consistent and $\|\left(1-\frac{\delta_i}{\hat{\delta_i}}\right)(\hat{\nu_i}-\nu_i)\|_2=\op(n^{-1/2})$, the proposed estimator $\hat{\bbeta}_{UW}$ is as efficient as the estimator derived using the true nuisance functions. In the Plugin method, the same property would require $\|\hat{\nu_i}-\nu_i\|_2=\op(n^{-1/2})$, which is even not achievable by typical parametric estimators. 

Theorem \ref{thm:high_better} asserts that the estimator $\hat{\bbeta}_{UW}$ is asymptotically more efficient than  $\hat{\bbeta}_{W}$.

\begin{theorem}\label{thm:high_better} 
    Assume that conditions in \Cref{thm:asymp_norm} holds for $\hat{\mu}$ and $\mu$ in places of $\hat{\nu}$ and $\nu$. Then, $\sqrt{n}(\hat{\bbeta}_{W}-\bbeta)\xrightarrow{D} \cN(0, \Sigma_{W})$ and $ \Sigma_{UW} \preccurlyeq \Sigma_{W}$. 
\end{theorem}

\begin{remark}
 The goal of this analysis is to identify peptides whose abundances are marginally associated with a given feature, such as disease status, cell types, or other phenotypic variables, which is a commonly studied problem in the field. Therefore, we adopt a regression model $Y^j=\bW^T\beta^j+\epsilon^j$ for peptide $j$, which doesn't include any auxiliary peptide data. For imputation, we use other peptide data as auxiliary variables because we want to approximate $Y^j$ with less variability, and we assume that the noise variables 
$\epsilon^1,\cdots,\epsilon^p$ are correlated between peptides. In this way, we examine the marginal association between the variables of interest while incorporating other peptide data outside the model without contradiction.

\end{remark}

\begin{remark}
The results presented in this section assume that the nuisance functions $\hat{\nu}$ and $\hat{\delta}$ are estimated from samples independent of $(Y_i, C_i, \bW_i, \bU_i)$. This assumption is used for the brevity of the presentation.
There are two standard approaches to improve the sample efficiency loss due to data splitting.  The first is 
cross-fitting \citep{chernozhukov2018double, kennedy2023towards}, which swaps the subsamples used for nuisance estimation and regression inference, and combines the test statistics from different folds to obtain the final inference. 
Alternatively, if the nuisance estimates belong to a Donsker class, then one can use empirical process theory to establish the asymptotic normality without sample splitting \citep[see Lemma 19.24 of][for example]{van2000asymptotic}.  Both approaches can be combined with the method proposed in this paper in a straightforward manner.
In our numerical experiments and data analyses, we used the same data for nuisance estimation and post-imputation OLS inference.  The good performance of our method suggests that the nuisance estimates in these settings are probably regular enough for the empirical process theory to work.
\end{remark}

\section{Multiple testing procedure for peptides}\label{sec:multiple}

The p-values derived in  \Cref{sec:method}, combined with a multiple testing procedure, allow us to make discoveries of important peptides associated with a covariate of interest. \Cref{subsec:algorithm} provides a detailed algorithm, and \Cref{sec:simulation} investigate its performance compared to the benchmark methods. The same algorithm is applied to real data studies in \Cref{sec:sc,sec:AD}.

\subsection{The input data and the algorithm} \label{subsec:algorithm}

For $n$ i.i.d. samples, the observed abundances of $p$ peptides can be written as an $n\times p$ matrix $\bC \odot \bY \in \mathbb{R}^{n \times p}$, where $\bC \in \{0,1\}^{n \times p}$ indicates the entry-wise missingness and $\bY\in\mathbb R^{n\times p}$ is the full data matrix without missing. Here ``$\odot$'' stands for the component-wise product. Only $\bC$ and $\bC\odot \bY$ are available.
Also observed is a covariate data matrix $\bW \in \mathbb{R}^{n \times q}$. Note that, for each observation $i$, the observed abundance of peptides $\{Y_{ij}:C_{ij}=1,j\in[p]\}$ provide extra information to impute the unobserved ones $\{Y_{ij}:C_{ij}=0,j\in[p]\}$. Consequently, we treat $\bU_{i}^j :=\{C_{i\ell} Y_{i\ell}:\ell\neq j\}$ as the augmented covariate for each peptide $j$.
The inference task is to test the significance of regression coefficients for the low dimensional covariates in $\bW$ on each peptide.  To this end, we will obtain individual p-values for each peptide using the asymptotic results presented in the previous section and then apply a multiple-testing framework such as the Benjamini-Hochberg procedure \citep{benjamini1995controlling}.

The procedure involves the estimation of two nuisance functions: the propensity score function $\delta$ and the augmented regression function $\nu$. In our upcoming experiments, we use Logistic regression to estimate $\delta$. The estimation of $\nu$ requires repeatedly regressing each column of $\bY$ on both the low-dimensional covariate $\bW$ and the other peptides as the high-dimensional auxiliary covariate $\bU$.  
An additional challenge is that each column of $\bY$ has many missing entries, even when used as a covariate in the regression problem, resulting in unregular auxiliary covariates for different samples. 
To address this issue, we use a nonparametric deep neural network model, VAEIT \citep[][see Appendixl B for details]{Du:2022}, which allows for flexible input and simultaneous estimation of the multi-response regression.  
More specifically, VAEIT utilizes masking strategies to model the mapping $\nu_j: \bW_i,\bU_{i}^j \mapsto Y_{ij}$ for $j\in[p]$. In other words, it learns regression functions $\nu_j(\bW_i,\bU_{i}^j) = \EE[Y_{ij}\mid \bW_i,\bU_{i}^j]$ for $j\in[p]$, enabling the simultaneous imputation of all missing peptides.
Although the convergence rates of estimating individual nuisance functions may be slower than $\sqrt{n}$, the doubly robust procedure allows for valid statistical inference as long as the product of the two convergence rates is $\op(n^{-1/2})$, as illustrated in \Cref{thm:asymp_norm}. Finally, the whole procedure is summarized in Algorithm \ref{algori}.

\begin{algorithm}
\caption{Multiple testing procedure for peptides}\label{algori}
\begin{algorithmic}
\Require Observed outcome $\bC \odot \bY \in \mathbb{R}^{n \times p}$; Observability $\bC \in \mathbb{R}^{n \times p}$; Covariates $\bW \in \mathbb{R}^{n \times q} $
\State Estimate $\hat{\bm{\nu}} \in \mathbb{R}^{n \times p}$ by running VAE on $(\bC,\bC\odot\bY,\bW)$.  
\For{$j=  1,\cdots,p$}
 \State Rewrite $\bY_i=(Y_i, \bU_i) \in \mathbb{R}^{1} \times \mathbb{R}^{p-1}$ where $Y_i=\bY_{ij}$, $\bU_i=\bY_{i(-j)}$ and $C_i=\bC_{ij}$.
 \State Estimate $\hat{\delta}_{i}$ by regressing $C_{1},\cdots,C_{n}$ on $\bW$ by logistic regression.
 \State Compute pseudo-outcomes $\hat{ Y}^{UW}_{i}=\frac{C_{i}}{\hat{\delta}_i} Y_{i}+(1-\frac{C_{i}}{\hat{\delta}_i}) \hat{\nu}_{ij} $. 
 \State Regress $\hat{Y}^{UW}_1,\cdots \hat{Y}^{UW}_n$ on $\bW_1,\cdots,\bW_n$ and compute a p-value ($P_j$) for the covariate of interest based on asymptotic distribution given in \Cref{thm:asymp_norm}.
\EndFor
\State \Return $P_1,\cdots,P_p$
\State Transform $P_1,\cdots,P_p$ to Benjamini-Hochberg's q-values and select indices whose q-values are less than a predefined cutoff. 
\end{algorithmic}
\end{algorithm}

To achieve the double robustness advantages, the proposed method requires consistent estimates of the nuisance parameters. The consistency of the  propensity score estimator $\hat{\delta}$ follows through theoretical arguments. The logistic model is nonlinear in parameters and some iterative procedures, such as Newton-type methods, are applied to find the MLE of parameters. When the number of parameters is far less than the sample size, then such procedures are known to achieve a global convergence \citep{gourieroux1981asymptotic, lee2014proximal}. \citet{rashid2009consistency} provide an extensive Monte Carlo simulation result to verify a finite sample performance. The consistency of $\hat{\nu}$ is more difficult to prove, and we provide an empirical result to show that the mean squared error consistently decreases as the sample size increases in the simulated dataset (Figure C1).

\subsection{Simulation study}\label{sec:simulation}

We investigate the performance of our method compared to several other methods on simulated data.  Eight methods are compared; Full, Complete, MICE, SVD, MissForest, DR\_W, DR\_UW (proposed), and Plugin, where they differ in the approach to obtain $P_1,\ldots,P_p$ in \Cref{algori}. The Full method uses the practically unavailable data $\bY$ without missingness. We perform a linear regression for each column of $\bY$ on low-dimensional covariates $\bW=(a,x)$, where $a$ is the variable of interest and $x$ represents any other covariates. We then use a linear regression $t$-test to decide if the coefficient of $a$ equals zero.  
The Complete method works in the same way as the Full method, but uses only observed samples.
The MICE method uses a multiple imputation method to impute missing values and then performs the test if the coefficient of $a$ is equal to zero using a statistic proposed by \cite{rubin2004multiple}. 

MICE is not computationally feasible when high-dimensional auxiliary variables are used for imputation. For this reason, missing values are imputed on the basis of low-dimensional covariates only. The SVD and MissForest methods use an approach we call ``Plugin-missing'', where only the missing values are imputed while the observed values remain unchanged. SVD estimates the missing values as a linear combination of the 10 most significant peptides \citep{troyanskaya2001missing}, and MissForest makes predictions iteratively based on random forests until convergence \citep{stekhoven2012missforest}. The DR\_W method is similar to \Cref{algori}, but the columns of $\hat{\bm{\nu}}$ are fitted by a linear regression model only with low-dimensional covariates. The DR\_UW method follows \Cref{algori}. The Plugin method regresses each column of the fitted outcomes $\hat{\bm{\nu}}$ in \Cref{algori} on the low-dimensional variables and performs a linear regression $t$-test for the coefficient of $a$. All seven methods, except for the MICE method, require a choice of variance estimator to perform the linear regression $t$-test. For the Full, Complete, Plugin-missing and Plugin methods, the usual OLS variance estimator is used. For the DR\_W and DR\_UW method, either the usual OLS variance estimator (in Models 1 and 2 below) or the heteroskedastic-consistent estimator \eqref{eq:hetero} (in Models 3 and 4) is used. The eight methods repeat the same procedure to obtain the p-values for each column of the outcome matrix. Then we transform the p-values into the Benjamini-Hochberg q-values and select the indices whose q-values are less than a predefined cutoff $\alpha$ to identify the discoveries. For each of the eight methods, the fraction of false discoveries over the number of total discoveries (FDR; False Discovery Rate) and the fraction of true discoveries over the number of signal peptides (TPR; True Positive Rate) are reported. An ideal method would control FDR within $\alpha$, and have a TPR close to one. Two sample sizes of $n=200, 500$ and a dimension $p=1000$ are considered. The number of repetitions is 200. 

\begin{remark}

The Plugin-missing approach lies between the Plugin and the Complete methods and is another widely used technique in practice, especially when the imputation algorithm only provides an estimate for the missing values. For the VAE result, we only consider a combination with the Plugin method because it is the standard approach when the imputation algorithm outputs estimates for the entire matrix. This is sometimes referred to as a denoising procedure; \cite{wang2019data} provides a comparative analysis of these procedures. 
\end{remark}

The simulation data are generated as follows. For the $j$th peptide and the $i$th sample, the outcome $y^j_i$ is formulated as $$y^j_i=\beta_{x,j} x_i+\beta_{a,j} a_i+\epsilon^j_i$$ for $j \in \{1,\cdots,p\}$ and $i \in \{1,\cdots,n\}$.  A case-control label $a_i$ is generated by selecting the $0.5n$ indices from $\{1,\cdots,n\}$ and setting $a_i=1$ for the cases. Otherwise, $a_i=0$ for controls. To introduce differential abundance, we randomly select $0.1p$ peptides and inject positive signals into the case data. We denote $s^j=1$ if $j$ is selected and call it a signal peptide; otherwise, $s^j=0$ and we call it a null peptide. A coefficient of interest $\beta_{a,j}$ is positive if $s^j=1$, and zero otherwise.

Four scenarios are considered, including missing patterns, Gaussian and skewed distributions of abundance data, and various forms of the true regression model: 
\begin{itemize}
    \item [Model 1.] Gaussian data without $X$ (MCAR); $y^j_i=c_1 s^j a_i+\epsilon^j_i$
    \item [Model 2.] Gaussian data (MCAR); $y^j_i=x_i+c_1 s^j a_i+\epsilon^j_i$        
    \item [Model 3.] Gaussian data (MAR); $y^j_i=x_i+c_1 s^j a_i+\epsilon^j_i$
    \item [Model 4.] Skewed data (MAR); $y^j_i=x_i+c_2 s^j a_i+\epsilon^j_i$
\end{itemize}

\noindent  Correlation between peptides is simulated using a realistic covariance structure to model the noise terms associated with each peptide; The covariance ($\Sigma$) was estimated from peptides measured in bulk brain tissue \citep{MacDonald:2017}. For Models 1, 2 and 3, we simulate $n$ i.i.d. vectors $(\epsilon_i^1,\ldots,\epsilon_i^p)$, $i=1,\ldots,n$, using a multivariate normal distribution with zero mean and covariance $\Sigma$. For Model 4, we generate skewed noise as follows: simulate multivariate normal variables as before, for each peptide add a constant to ensure that all entries are positive, apply a log transformation, and finally recenter each peptide at zero.
Covariates $x_1,\cdots,x_n \in \mathbb{R}$ are generated independently from a uniform distribution in $(0,1)$. After generating covariates and noise, each outcome $y^j_i$ is randomly masked with the probability determined from the missingness model. In Models 1 and 2, each $y^j_i$ is missing completely at random (MCAR) with equal probability: $\PP(C_{ij}=0)=0.3$. In Models 3 and 4, $y^j_i$ is missing at random (MAR): $\PP(C_{ij}=0)={e^{x_{i}}}/\{2(1+e^{x_{i}})\}$. We use different signal strengths for different sample sizes to construct meaningful comparisons between methods; $c_1=0.4$, $c_2=0.12$ for $n=200$ and $c_1=0.3$, $c_2=0.08$ for $n=500$.

\begin{figure}[!ht]
\centering
	\includegraphics[width=0.8\textwidth]{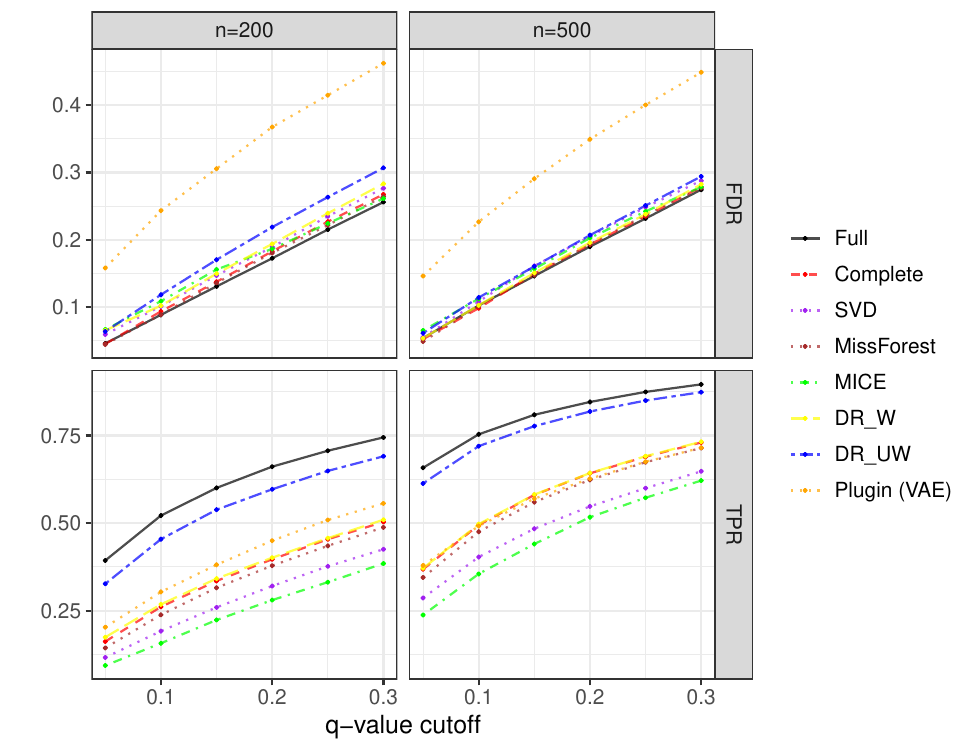}
\caption{ Performance of different methods on simulated data according to Model 3. The X-axis represents different levels of q-value cutoffs, and the Y-axis represents either FDR or TPR. }
\label{fig:sim_result3}
\end{figure}

Figure \ref{fig:sim_result3} summarizes the FDR and TPR of eight methods applied to Model 3. As expected, FDR is well controlled for the Full, Complete, MICE, DR\_W, and DR\_UW methods, 
whereas the Plugin method inflates the FDR. This occurs because the differences between the case and control data in signal peptides bleed into correlated null peptides during the imputation procedure. When no signals are injected into any of the peptides or when the signal is carried for sets of correlated peptides, the Plugin method is also well-controlled (results not shown). 

Naturally, the Full method demonstrates the highest TPR, representing the optimal performance achievable in this setting without missing data. DR\_UW shows the second-best TPR and it becomes similar to the Full method when $n$=500. Complete, MICE, DR\_W, and Plugin attain smaller TPR, with MICE being the most conservative. For the Complete method, this is expected because it excludes samples with missing data.  MICE and DR\_W perform an imputation based on low-dimensional variables, which is less accurate, leading to greater variance. 
Models 1,2 and 4 produce similar results (see Appendix C). 

Two Plugin-missing methods control the FDR, but the TPR is even lower than that of the Complete method, indicating that there is no benefits from the additional simulation step. When stronger signals are injected (i.e., $y^j_i=x_i +s^j a_i+\epsilon^j_i$ in Model 3), they have inflated FDR (Figure C4). Thus, the Plugin-missing method performs somewhat between the Plugin method and the Complete method, which is not surprising, as the response in the Plugin-missing methods only uses a partial substitution of the original data.

\section{Single-cell protein abundance varies with cell size}\label{sec:sc}

 For illustration, we analyze a single-cell proteomic dataset based on mass spectrometry as published by \citet{Leduc:2022}. Data processing, as described by the authors, involves several steps, including QC and a log transformation.  The primary interest is detecting peptides whose abundance varies strongly with cell size. A peptide is a short chain of amino acids that constitute proteins and is a useful unit for quantitative analysis. 
The proportion of observed cells differs significantly between the peptides, and for 85.6\% of the peptides, the observation rate is smaller than 0.5.  When the observation rate is too low, it is not reasonable to expect that any method will perform satisfactorily; therefore, we focus on peptides with a missingness of no more than 50\%. Because the threshold for the observation rate is controversial, for the main analysis, we provide a range of results with respect to different thresholds (0.5, 0.6, 0.7, 0.9). For other parts of the analysis, including exploratory data analysis and realistic simulation, we focus on a threshold of 0.7. After removing peptides whose observation rates are less than 0.7, there are a total of 753 remaining peptides.

\begin{figure}[h]
\centering
	\includegraphics[width=13cm]{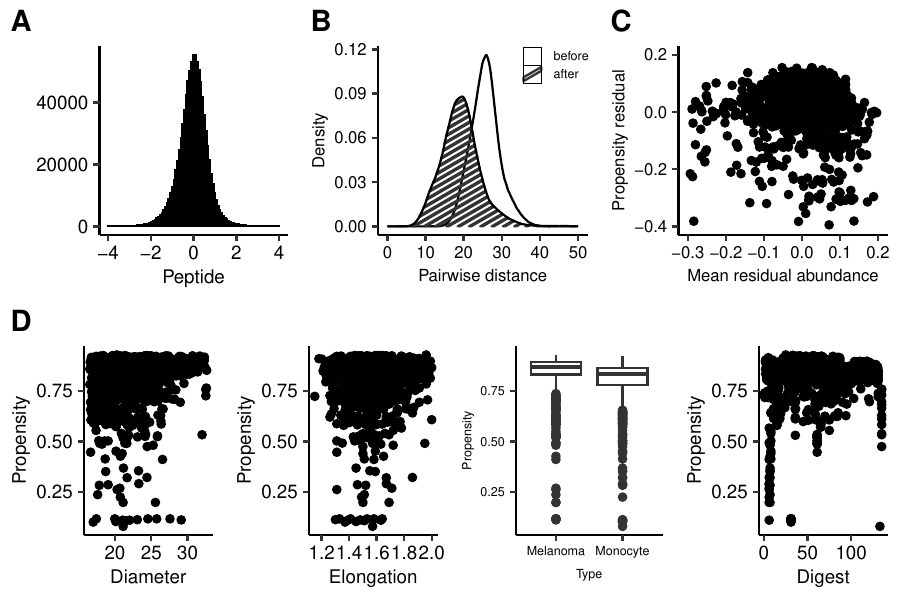}
\caption{ (A) Histogram of peptide abundance data (B) Distribution of pairwise distance between cells before and after an imputation (C) Scatter plot between propensity residual and Mean residual abundance (D) Scatterplot between Propensity score and other cell-level covariates.} 
\label{fig:FigureC}
\end{figure}

We first present exploratory data analysis to provide a rational basis for applying our method. The distribution of cell-wise peptide abundance data for peptides with more than 70\% of observed rates reveals a Gaussian-like distribution (Fig. \ref{fig:FigureC}A); thus, these data are well suited to our imputation model, which is a VAE model tailored to a Gaussian distribution; see Appendix B for more details. The distribution of pairwise distances between cells before and after imputation shows a noticeable reduction in distances after imputation, indicating that the overall variance of abundance data has decreased (Fig. \ref{fig:FigureC}B). Next, we check the assumption of MAR by examining the relationship between the measured variables and the observed cell-wise rate among the peptides (propensity score). Four measured variables are examined: cell diameter, elongation, type, and digest. The lack of a relationship between the residuals of the estimated propensity score and the mean abundance in cells, after regressing each of them into four covariates, shows that the observed relationship between the abundance of the peptide and the propensity score is largely explained by the measured covariates (Fig. \ref{fig:FigureC}C). However, the joint distribution of the cell-specific propensity score and each covariate, along with its marginal distribution, illustrates that each covariate is related to the propensity score to some extent, supporting an analysis under the assumption of MAR (Fig. \ref{fig:FigureC}D). Furthermore, a reasonable imputation model can be built upon the robust relationship between peptides. Examining the quantile value of 0.9 of the absolute correlation coefficient for each peptide with other peptides reveals a strong correlation pattern. These values generally fall between 0.1 and 0.5, providing a good foundation for a high-dimensional imputation model (see Figure C6 for details).

This section is organized as follows. First, we conduct realistic simulations to see how the proposed method works on this dataset with an artificially generated ground truth.
Next, we present the results for the main analysis, where the primary focus of our analysis is to identify peptides whose abundance varies with the diameter of the cell. For all settings, we compare the proposed DR\_UW method with the Complete, DR\_W, and Plugin methods. Naturally, the Full method cannot be considered.

\subsection{Realistic simulation} Before going into the main analysis, we check the performance of the proposed method with some artificially generated ground truth incorporated into this dataset and compare different methods in terms of TPR and FDR. Specifically, we artificially generate the type variable (case and control) while keeping all other variables (cell diameter, elongation and digest) unchanged. In Setting 1, we randomly permute the measured type variable among the cells. In Setting 2, the type variable is randomly generated from a Bernoulli distribution with a probability proportional to the propensity score. 
 Signal peptides are randomly selected for 10\% of the total considered peptides, and then a positive signal generated from a normal distribution with mean 0.2 and variance 0.05 is added to the case cells in the signal peptides. After imputation, we identify differentially abundant peptides using various q-value cutoffs of 0.01, 0.05, and 0.3.

 The Complete, DR\_W, and DR\_UW methods detect a reasonable proportion of signal peptides while maintaining control of FDR in both settings.  For FDR=0.01 and 0.05, the DR\_UW method provides better TPR than the Complete and DR\_W method. The Complete method is better than the DR\_W method in TPR, indicating that low-dimensional imputation is too noisy for satisfactory results. For FDR=0.3, all three methods, Complete, DR\_W, and DR\_UW, achieve near-perfect TPR. The Plugin method severely inflates FDR for all FDR levels, and the TPR is lower than the other three methods. These results are summarized in Figure C7 and C8 in Appendix.   Notably, the improvement in results is not as striking as those displayed in Figure \ref{fig:sim_result3}, which are based on bulk tissue simulations.  The value of the proposed method depends on the quality of imputation we can obtain from a dataset.  As single-cell data are more difficult to impute due to a higher noise level and greater missingness, it is not surprising that the performance is somewhat diminished.

\subsection{Main analysis} 
\begin{figure}[h]
\centering
	\includegraphics[width=10cm]{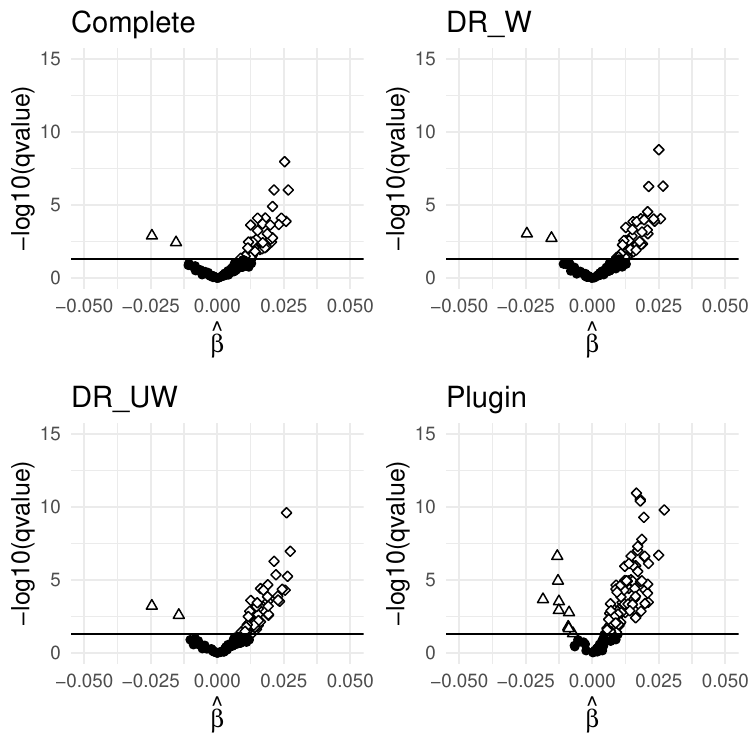}
\caption{Volcano plot of peptide discoveries by different methods in the single-cell proteomics dataset analyzed in \Cref{sec:sc} when an observability threshold of 0.9 is applied. The horizontal line indicates the q-value cutoff of 0.05.} 
\label{fig:fourplots}
\end{figure}

One objective of this analysis is to detect peptides whose abundance varies strongly with cell size. We first filter the peptides by applying varying thresholds (0.5, 0.6, 0.7, 0.9) to the observation rates of the peptides and focus on analyzing those peptides.  A larger proportion of peptides, whose observed rates are greater than 0.2, is used to feed the imputation procedure. After imputation, peptides are selected based on linear regression models: $$\text{Peptide abundance} \sim \text{Diameter+Type+Digest+Elongation}$$ where we compute the p-values associated with the diameter variable. The p-values are transformed into q-values using the BH procedure.  Based on estimated coefficients and the corresponding q-values, the Complete, DR\_W, and DR\_UW methods exhibit roughly similar distribution patterns; however, the Plugin method yields a larger number of significant q-values compared to the other three methods due to the signal bleeding effect (Fig. \ref{fig:fourplots}).  

 The objective of this analysis is to detect peptides whose abundance varies strongly with cell size. The impact of cell size on cell physiology is of interest in two domains: large cell size may be a cause rather than a consequence of cell senescence \citep{lanz2022increasing,jones2023characterization}; and cell size is a determinant of stem cell fate \citep{lengefeld2021cell}.
Cells develop various shapes and sizes to perform their specific functions and the coordination of cell growth and division to achieve and maintain these cellular phenotypes has been a key research interest in cell biology \citep{turner2012cell, lloyd2013regulation, liu2024cell}.  Because large cells need to produce and maintain more cellular components than small cells, it is anticipated that cell size will be positively associated with some protein abundances, and this is numerically validated in multiple earlier studies \citep{Marguerat:2012,lanz2022increasing, jones2023characterization};  
however, there could be exceptions.  For instance, a small cell that is growing rapidly might produce more of the proteins that are essential for this current developmental phase, and this phenomenon would lead to a small number of negative associations. Regardless of the direction, identifying the proteins associated with cell size could provide insights into abnormal cell growth, such as in cancer or stem cell fate \citep{lengefeld2021cell, jones2023characterization}.

 If the majority of true signals are positive, it follows that the estimated $\beta$ coefficients will tend to be positive for the signal peptides and symmetrically distributed around zero for the null peptides. Therefore, most of the significant coefficients will be positive, and only a small number of noisy null peptides will contribute to both positive and negative signs of the selected coefficients. We propose the \textit{mirror rate}, defined as the number of significant peptides with $\hat{\bbeta}<0$ divided by the number of significant peptides with $\hat{\bbeta}>0$, as a metric to compare the reliability of the findings. As more noisy null peptides are included, the mirror rate will increase. 
Several recent papers \citep{dai2023scale, guo2023threshold, du2023false} derive an FDR metric from a statistic that exhibits such an asymmetric structure of null and non-null (signal) scenarios. In practice, we anticipate that our mirror rate may overestimate the false discovery rate due to the possibility of some proteins having a negative relationship. However, since such a fraction of relationships are commonly present in all the methods we compare, we expect it to provide a useful comparison across methods.

The performance of each method is evaluated in settings with different thresholds for observation rate and q-values (Table \ref{table:singlecell}). 
We first fix the q-value cutoff at 0.05 and apply different thresholds to the observation rates of peptides. For a threshold of 0.9, mirror rates are well-controlled for the Complete, DR\_W and DR\_UW methods, but the number of discoveries is relatively small because many peptides are excluded from the analysis. The Plugin method provides the largest discoveries, but its mirror rate is inflated. When the threshold is lowered to 0.7, 0.6, and 0.5, the number of discoveries becomes larger, and mirror rate tends to increase. This is natural because if the peptides with a high rate of missingness are introduced, the inference problem becomes more challenging. However, compared to the Plugin method, the other three methods consistently give better control of the mirror rate, the proposed method DR\_UW being the best. In addition, DR\_UW provides a larger number of discoveries. This is consistent with what we observed from the simulations. It controls the FDR well while achieving greater TPR.  The Plugin method provides the largest number of discoveries, but it generally inflates the mirror rate. Similar results hold when we fix the threshold to 0.7 and apply different q-value cutoffs.

\begin{table}[h] 
\begin{tabular}{c c c c c c c c c c}
\toprule
\multirow{2}{1.5cm}{\centering Observability threshold} & \multirow{2}{1cm}{\centering q-value cutoff} &\multicolumn{4}{c}{Mirror rate} & \multicolumn{4}{c}{Number of selected peptides}\\ \cmidrule(lr){3-6} \cmidrule(lr){7-10}
    && {Com}  & {DR\_W} & {DR\_UW} & Plugin & {Com} & {DR\_W} & {DR\_UW} & Plugin \\ \cmidrule(lr){1-2} \cmidrule(lr){3-6} \cmidrule(lr){7-10}
    0.9    &\multirow{4}{*}{0.05}  & {0.05} & {0.05}  & {0.04}   & 0.10    & {40}  & {44}    & {51}  & 86      \\ 
    0.7       &    & {0.20} & {0.22}  & {0.13}   & 0.26    & {111} & {106}    & {149}    & 303     \\ 
    0.6      &                    & {0.31} & {0.33}  & {0.25}   & 0.40    & {133} & {128}   & {186}    & 419     \\ 
    0.5      &                    & {0.41} & {0.45}  & {0.35}   & 0.55    & {158} & {152}   & {218}    & 535    \\   \hline \rule{0pt}{3ex}    
\multirow{4}{*}{0.7}&0.01   & {0.20} & {0.25}  & {0.15}   & 0.20    & {46}  & {44}    & {73}     & 225     \\ 
&0.05    & {0.20} & {0.22}  & {0.13}   & 0.26    & {111} & {106}    & {149}    & 303     \\ 
&0.1   & {0.20} & {0.22 }  & {0.19}   & 0.28    & {148} & {144}   & {189}    & 334     \\ 
&0.3    & {0.27} & {0.31}  & {0.29}   & 0.33    & {267} & {258}   & {304}    & 390     \\ \bottomrule
\end{tabular}
\caption{Mirror rate and number of peptides selected with each method under different combinations of thresholds applied to an observed rate of peptides and q-value cutoffs.}
\label{table:singlecell}
\end{table}

To further verify the robustness of the result, we map the peptides discovered by DR\_W, DR\_UW, and the Plugin method to the corresponding proteins, and check their overlaps with the discoveries of the Complete method (a q-value cutoff 0.05) at the protein level. We assess their contributions by only considering additional discoveries beyond those made by the Complete method with a q-value cutoff of 0.01. A threshold of 0.7 is applied to the observation rate of peptides. When applying a q-value of 0.01, the additional discoveries of the DR\_UW method are largely robust at the protein level; 90\% of them overlap with those discovered by the Complete method. As we increase the q-value cutoff to 0.05, 0.1, and 0.3, the proportion of such overlaps tends to decrease, but the number of additional discoveries is much larger. Across all q-value cutoffs, the DR\_UW method consistently provides more additional discoveries than the DR\_W method, and has a slightly smaller proportion of overlaps with the Complete method. This aligns with theory and simulation results in the sense that both doubly robust methods exhibit reasonable control of the false discovery rate, while DR\_UW demonstrates better efficiency. The Plugin method discovers the largest number of additional peptides, but as expected, its findings do not largely overlap with the discoveries of the Complete method at the protein level. The detailed results are summarized in Table C1.

\section{Peptide abundance in key proteins is associated with Alzheimer's disease}\label{sec:AD}

Alzheimer's disease (AD) is a prominent neurodegenerative disorder among older adults. Numerous environmental and genetic factors are known to contribute to the disease, and related biological pathways have yet to be fully discovered. In this section, we apply the proposed method to identify important peptides associated with AD and related dementias. A bulk peptide-level dataset offers an opportunity to illustrate these methods in an important scientific setting \citep{Merrihew:2023}. While samples in this dataset are annotated with a range of disease severity, we group them into two types; cases (samples with autosomal dominant/sporadic AD dementia) and controls (samples without dementia, with or without a high AD histopathologic burden).  Two other covariates, the brain region, and PMI, are also used in our analysis.

In this analysis, we focus on peptides whose observed rates are between 0.5 and 1 in each of the four brain regions. If the observed rate is 1, the Complete method and the DR methods will provide the same selection result. The propensity scores for each sample are mostly around 0.9, and we assume the MCAR missing pattern. Some of the samples have missing covariates on PMI. After further removal of these samples, we have 488 peptides and 220 samples, including 139 cases and 81 controls. The abundance distributions of the peptides vary significantly between the brain regions. Therefore, we apply the VAE model separately to each brain region. Although VAE is usually applied to a dataset with a large sample size, when data have a Gaussian-like distribution, as in peptides, VAE gives a robust imputation outcome. The final selection of peptides is based on linear regression models: $$\text{Peptide abundance} \sim \text{Type+Region+PMI}$$ where we compute p-values associated with the type variable. BH procedure is then applied to convert them into q-values. The final discoveries are determined by applying a cutoff of 0.05 to them.  Following this procedure, the Complete method selects 55, DR\_W selects 50, DR\_UW selects 58, and Plugin selects 79 peptides. The Complete, DR\_W, and DR\_UW methods have a similar number of discoveries. Only peptides with a low missingness rate are shared for these data. Hence, there is limited opportunity for improvements over the complete method.  While the Plugin method selects more peptides than the other methods, it is not surprising that the number of discoveries does not vary much across the well-calibrated methods. 

Seven peptides are selected by the DR\_UW method but not by the Complete method (Table \ref{table:AD}).  To determine whether these discoveries are meaningful, we examine the corresponding protein and gene annotations. Six have been linked to AD and related literature (reference papers are listed in the final column of the table). Specifically, the genes PDE2A, NEUM, MX1, and CGT have revealed a direct connection with AD in the corresponding reference papers. The ANK2 gene is associated with autism and the reference paper links Drosophila ANK2 (human ANK1) to the characteristics of AD. The protein sp|Q9H305|, associated with the CDIP1 gene, plays a major role in controlling cell death, a feature of AD, and the gene is highly expressed in the brain, which implies a possible connection to AD at some level. The peptide CALD1 is only found to have an indirect connection to AD. It belongs to a group of pathway genes that change with age and are reversed by Riluzole, which is related to synaptic transmission and plasticity.

\begin{table}[h]
\begin{tabular}{ccc R{5.5cm}}
\toprule
\textbf{Peptide} & \textbf{Protein} & \textbf{Gene} & \textbf{Reference}\\ \midrule
MPLYGLHLWLPK & sp|P03905| & NU4M & \cite{bhatia2022mitochondrial, wesseling2017system}      \\\addlinespace[0.5ex]
NLFTHLDDVSVLLQEIITEAR & sp|O00408| & PDE2A & \cite{sheng2022inhibition, Shi:2021tp, Delhaye:2024wf}      \\\addlinespace[0.5ex]

TTHRPHPAASPSLK & sp|Q01484| & ANK2 & \cite{kumari2022temporal, Higham:2019vq} \\\addlinespace[0.5ex]
MQNDTAENETTEKEEK & sp|Q05682| &CALD1 & \cite{pereira2017age} \\\addlinespace[0.5ex]
LGVSFLVLPK & sp|Q16880| & CGT &         \cite{Tang:2023tz, Moll:2020vl, ryckman2020metabolism} \\\addlinespace[0.5ex]
NFEEFFNLHR & sp|P20591| & MX1 & \cite{Prakash:2024tf, Widjaya:2023ve, ma2012mxa} \\\addlinespace[0.5ex]
DVTHTC{[}+57{]}PSC{[}+57{]}K & sp|Q9H305| & CDIP1 & \cite{dileep2023neuronal, Inukai:2021ui} \\ \bottomrule
\end{tabular}
\caption{\rm Peptides discovered by DR\_UW and not by the Complete method. A q-value cutoff of 0.05 is used.}
\label{table:AD}
\end{table}

\section{Discussion} \label{sec:discussion}

In this paper, we present a statistical framework for analyzing proteomic data with missing values. Our proposed estimator $\hat{\bbeta}_{UW}$ is established in a doubly robust framework and achieves reduced asymptotic variance leveraging correlations between different peptides.  
Through simulations and real data analysis, we demonstrate that the proposed estimator offers highly competitive statistical decisions in discovering signal peptides.

The doubly robust method requires that the estimated conditional mean, $\hat{\nu}$,  converges to ${\nu}$; however, it allows for this convergence to be slow—just not too slow.  By contrast, under such circumstances, the natural competing imputation approach, the Plugin method, results in invalid inference (inflated FDR). 
At the performance extremes, there is nothing to gain: if $\hat{\nu}$  converges rapidly, then every imputation method works; and if $\hat{\nu}$ is inconsistent or converges very slowly, then none of the competing imputation methods are valid. 
This Goldilocks setting has two important implications: first, in all scenarios, the DR method never harms us, and there is a regime where it provides significant benefits; and second, we can roughly check the quality of the imputation through simulation results to determine the validity and efficiency of the DR approach.t`

One might question why the doubly robust procedure offers significant advantages over simply replacing the missing data with the fitted mean value.  Formally, let us consider two alternative estimators for the response $g(\cdot; \nu, \delta)$: plugin-missing, $\nu(\bW_i,\bU_i)+C_i(Y_i-\nu(\bW_i,\bU_i))$, and doubly robust, $\nu(\bW_i,\bU_i)+\frac{C_i}{\delta(\bW_i)}(Y_i-\nu(\bW_i,\bU_i))$.  The plugin-missing estimator replaces $Y$ with our best estimate of $\nu$ only when the data are missing.  For the doubly robust estimator, the first term also represents the best estimate of $\nu$, while the second term is designed to minimize the impact of the estimation error. Intuitively speaking, the $1/\delta$ term gives greater weight to the residuals in the portion of the data where we rarely observe $Y$. This adjustment has the effect of removing the first-order error in the estimation of $\nu$, as shown in \Cref{eq:bias_2nd}. This property was derived from the function $g$ satisfying $\EE[\frac{\partial g(\cdot; \hat{\nu}, \hat{\delta}) }{\partial \hat{\nu}}]=\EE[\frac{\partial g(\cdot; \hat{\nu}, \hat{\delta}) }{\partial \hat{\delta}}]=0$. 

In simulations, we show that the proposed DR method possesses improved properties compared to other imputation-based methods, such as the Plugin, Plugin-missing and DR\_W methods. However, a final choice between imputation-based methods and the Complete method, which uses only observed values, should depend on the quality of imputation achievable. Simulations and real data considered in Sections \ref{sec:sc} and \ref{sec:AD} provide a rich foundation for high-quality imputation with the VAE model. These data have Gaussian-like distributions, sufficient sample size, and robust correlation structure between peptides. If these conditions are not sufficiently satisfied, even the DR\_UW method will not perform as well as the Complete method. For example, a simulation experiment reveals that if the outcome model in Section \ref{sec:sc} produces a completely noisy imputation, then the doubly robust method will yield fewer discoveries compared to the complete method (Figure C9). 
However, even in such cases, the estimate $\hat{\bbeta}_{UW}$ obtained by the DR method is similar to the estimate obtained by the Complete method (Figure C10). This follows because the consistency of $\hat{\bbeta}_{UW}$ is still guaranteed by the doubly robust property, provided the estimated propensity score is consistent. 

The value of the proposed method depends on the quality of the imputation procedure.  Although the DR approach is applicable regardless of the choice of imputation algorithm, we chose a refined VAE procedure that uses masking to robustly handle missing values as an integrated part of the procedure \citep{Du:2022}.  In the proteomic literature, one of the most commonly applied imputation procedures is a version of $k$-nearest neighbors (kNN).   The standard kNN procedure in use involves imputing the missing values based on the mean of the $k$ closest peptides \citep{harris2023evaluating}.  Close peptides are used instead of close samples because it is very difficult to define close samples when there is excessive missingness.  As a consequence, this kNN approach does not utilize low-dimensional covariates $\mathbf{W}_i$ in imputation and is potentially biased with fewer discoveries.  An alternative approach is to adopt a two-step method: initially, we impute missing values based on the closest peptides, and then, once the missing entries are filled, we impute the entire dataset based on the closest samples. When we apply this two-step approach to the AD dataset in \Cref{sec:AD}, DR\_W selects 65, DR\_UW selects 51, and the Plugin method selects 121. Although we cannot reach an exact conclusion, the result of the VAE model in \Cref{sec:AD} is more aligned with theoretical expectations: the DR\_UW method provides more discoveries than the DR\_W method.

\begin{acks}[Acknowledgments]
We thank the Editor and two anonymous reviewers for their helpful comments, Bernie Devlin for his assistance in contextualizing the findings in \Cref{sec:AD} within the scientific literature, and Chan Park for his help in identifying important reference papers. Haeun Moon is also affiliated with the School of Transdisciplinary Innovations (Integrative Data Science Program) and the Institute for Data Innovation in Science at Seoul National University.
\end{acks}

\begin{funding}
This project was funded by National Institute of Mental Health (NIMH) grant R01MH123184, NSF DMS-2015492 and DMS-2310764.
\end{funding}
 \bibliographystyle{asa}

\pagebreak

\appendix
\counterwithin{theorem}{section}
\setcounter{table}{0}
\renewcommand{\thetable}{\thesection\arabic{table}}
\setcounter{figure}{0} 
\renewcommand\thefigure{\thesection\arabic{figure}}
\renewcommand{\thealgorithm}{\thesection\arabic{algorithm}}

\section{Proof}



\subsection{Proof of \Cref{thm:DR}} 

\begin{proof}
By plugging in $\hat{Y}_i^{UW}= \frac{C_i}{\hat{\delta}_i}Y_i+\left(1-\frac{C_i}{\hat{\delta}_i}\right)\hat{\nu}_i$, we have
\begin{align} 
\hat{\bbeta}-\bbeta&=\left(\sum_{i=1}^n \bW_i\bW^T_i\right)^{-1} (\sum_{i=1}^n \bW_i (\hat{Y}_i^{UW}-\bW^T_i\bbeta)) \nonumber\\
&=(\frac{1}{n}\sum_{i=1}^n \bW_i\bW^T_i)^{-1} (\frac{1}{n}\sum_{i=1}^n \bW_i (\epsilon_i+(1-\frac{C_i}{\hat{\delta_i}})(\hat{\nu_i}-\nu_i+Y_i-\nu_i)) \nonumber\\
&=(\frac{1}{n}\sum_{i=1}^n \bW_i\bW^T_i)^{-1} (\frac{1}{n}\sum_{i=1}^n \bW_i (\epsilon_i+(1-\frac{C_i}{\hat{\delta_i}})(Y_i-\nu_i) \nonumber \\ &\quad\quad\quad\quad\quad\quad\quad+(1-\frac{C_i}{{\delta_i}})(\hat{\nu_i}-\nu_i)+C_i(\frac{1}{\delta_i}-\frac{1}{\hat{\delta_i}})(\hat{\nu_i}-\nu_i))
\label{eq:100}
\end{align}

By assumption, $\frac{1}{n}\sum_{i=1}^{n}\bW_i\bW_i^T$ is nonsingular with probability approaching one, so its inverse exists. Let $D=\frac{1}{n}\sum_{i=1}^{n}\bW_i\bW_i^T-\EE[\bW\bW^T]$ which is $\op(1)$ by the law of large number.  Then, \begin{align}\label{eqn:5}
(\frac{1}{n}\sum_{i=1}^{n}\bW_i\bW_i^T)^{-1}&=(\EE[\bW\bW^T]+D)^{-1}\nonumber\\
&=\EE[\bW\bW^T]^{-1}(I+\EE[\bW\bW^T]^{-1}D)^{-1}\nonumber\\
&=\EE[\bW\bW^T]^{-1} (I-\EE[\bW\bW^T]^{-1}D(I+\EE[\bW\bW^T]^{-1}D)^{-1})\nonumber\\
&=\EE[\bW\bW^T]^{-1}+\op(1)
\end{align}
For the third step of Equation \ref{eqn:5}, we used $(M_1+M_2)^{-1}=M_1^{-1}-M_1^{-1}M_2(M_1+M_2)^{-1}$ proved in \cite{henderson1981deriving}. Therefore \begin{equation}
(\frac{1}{n}\sum_{i=1}^{n}\bW_i\bW_i^T)^{-1} \xrightarrow{p} \EE[\bW\bW^T]^{-1} \label{eq:101} \end{equation}

Also, we have
\begin{align*}
\mathbb{E}&(\bW_i (\epsilon_i+(1-\frac{C_i}{\hat{\delta_i}})(Y_i-\nu_i)+(1-\frac{C_i}{{\delta_i}})(\hat{\nu}_i-\nu_i))  ) \nonumber\\&=\mathbb{E}(\bW_i \mathbb{E}(\epsilon_i+(1-\frac{C_i}{\hat{\delta_i}})(Y_i-\nu_i)+(1-\frac{C_i}{{\delta_i}})(\hat{\nu_i}-\nu_i)|\bW_i,\bU_i)) \nonumber
\\&=\mathbb{E}(\bW_i (\mathbb{E}(1-\frac{C_i}{\hat{\delta_i}}|\bW_i,\bU_i)\mathbb{E}(Y_i-\nu_i|\bW_i,\bU_i)+\mathbb{E}(1-\frac{C_i}{{\delta_i}}|\bW_i,\bU_i)\mathbb{E}(\hat{v}_i-\nu_i|\bW_i,\bU_i))) \nonumber\\
&=0,
\end{align*}
where we used $\mathbb{E}(\epsilon_i|\bW_i,\bU_i)$, $\mathbb{E}(Y_i-\nu_i|\bW_i,\bU_i)=0$ and $\mathbb{E}(1-\frac{C_i}{\delta_i}|\bW_i,\bU_i)$ in a last step.

Then, by a weak law of large number, we have \begin{align}
\frac{1}{n}\sum_{i=1}^n \bW_i &(\epsilon_i+(1-\frac{C_i}{\hat{\delta_i}})(Y_i-\nu_i)+(1-\frac{C_i}{{\delta_i}})(\hat{\nu_i}-\nu_i)) \nonumber \\&\xrightarrow{P} \mathbb{E}(\bW_i (\epsilon_i+(1-\frac{C_i}{\hat{\delta_i}})(Y_i-\nu_i)+(1-\frac{C_i}{{\delta_i}})(\hat{\nu}_i-\nu_i))  )\nonumber \\&=0
\label{eq:102}
\end{align}   Also, for each $W_{ij}$ which denotes $j$th entry of a vector $\bW_{i}$, below inequality is obtained by using a Cauchy-Schwarz inequality
\begin{align}
\frac{1}{n}\sum_{i=1}^{n} W_{ij}C_i(\frac{1}{\delta_i}-\frac{1}{\hat{\delta_i}})(\hat{\nu_i}-\nu_i)  &\le \sqrt{\frac{1}{n} \sum_{i=1}^{n}W_{ij}^2}\sqrt{\frac{1}{n} \sum_{i=1}^{n}(\frac{1}{\delta}-\frac{1}{\hat{\delta_i}})^2(\hat{\nu_i}-\nu_i)^2} \nonumber \\
&\le \sqrt{\| W_{ij}\|_{\infty}^2}\sqrt{\|(\frac{1}{\delta}-\frac{1}{\hat{\delta_i}})(\hat{\nu_i}-\nu_i)\|_2^2} \nonumber \\
&=\op(1)
\label{eq:103}
\end{align}

Plugging in Equation  \ref{eq:101}, \ref{eq:102} and \ref{eq:103} to Equation \ref{eq:100}, we have a desired result.
\end{proof}

\subsection{Proof of \Cref{thm:asymp_norm}} 


\begin{proof}

 {\bf Part 1: limiting distribution of $\bbeta$}
  
By plugging in the formula for $\hat{\bbeta}$, we have \begin{align*}
	\hat{\bbeta}-\bbeta&=\left(\sum_{i=1}^n \bW_i\bW^T_i\right)^{-1}\sum_{i=1}^n \bW_i(Y_i+\left(\frac{C_i}{\hat{\delta}_i}-1\right)(Y_i-\hat{\nu}_i))-\bbeta\\
	&=\left(\sum_{i=1}^n \bW_i\bW^T_i\right)^{-1}\sum_{i=1}^n \bW_i(\epsilon_i+(\frac{C_i}{\hat{\delta_i}}-1)(Y_i-\hat{\nu}_i))
\end{align*}

Then $\sqrt{n}$-scaled difference is expressed as \begin{align}
	\sqrt{n}(\hat{\bbeta}-\bbeta)&=(\frac{1}{n}\sum_{i=1}^{n}\bW_i\bW_i^T)^{-1}\frac{1}{\sqrt{n}}\sum_{i=1}^{n}\bW_i(\epsilon_i+(\frac{C_i}{\hat{\delta_i}}-1)(y_i-\hat{\nu}_i)) \label{part0}
\end{align} 


As proved in Equation \ref{eq:101}, we have \begin{align}
    (\frac{1}{n}\sum_{i=1}^n \bW_i\bW^T_i)^{-1} \xrightarrow{P} \EE[\bW_i\bW_i^T]^{-1}
    \label{part1}
\end{align}

On the other hand, we can decompose \begin{align*}
(\frac{C_i}{\hat{\delta_i}}-1)&(Y_i-\hat{\nu_i})-(\frac{C_i}{{\delta_i}}-1)(Y_i-\nu_i)\\
&=(\frac{C_i}{\hat{\delta_i}}-1)(Y_i-\hat{\nu_i})-(\frac{C_i}{\hat{\delta_i}}-1)(Y_i-{\nu_i})+(\frac{C_i}{\hat{\delta_i}}-1)(Y_i-{\nu_i})-(\frac{C_i}{{\delta_i}}-1)(Y_i-\nu_i)\\
&=(\frac{C_i}{\hat{\delta_i}}-1)(\nu_i-\hat{\nu_i})+C_i(\frac{1}{\hat{\delta_i}}-\frac{1}{\delta_i})(Y_i-\nu_i)\\
&=(\frac{C_i}{\hat{\delta_i}}-\frac{C_i}{\delta_i})(\nu_i-\hat{\nu_i})+(\frac{C_i}{{\delta_i}}-1)(\nu_i-\hat{\nu}_i)+C_i(\frac{1}{\hat{\delta_i}}-\frac{1}{{\delta_i}})(Y_i-\nu_i)
\end{align*}
which leads to the equation below 
\begin{align}
\label{eqn:9}
\frac{1}{\sqrt{n}}\sum_{i=1}^{n}\bW_i(\frac{C_i}{\hat{\delta_i}}-1)(Y_i-\hat{\nu_i})&
=\frac{1}{\sqrt{n}}\sum_{i=1}^{n}\bW_i(\frac{C_i}{\hat{\delta_i}}-\frac{C_i}{\delta_i})(\nu_i-\hat{\nu_i})\\ \nonumber
&+\frac{1}{\sqrt{n}}\sum_{i=1}^{n}\bW_i(\frac{C_i}{{\delta_i}}-1)(\nu_i-\hat{\nu}_i)+\frac{1}{\sqrt{n}}\sum_{i=1}^{n}\bW_iC_i(\frac{1}{\hat{\delta_i}}-\frac{1}{{\delta_i}})(Y_i-\nu_i) \\ \nonumber
&+\frac{1}{\sqrt{n}}\sum_{i=1}^{n}\bW_i(\frac{C_i}{{\delta_i}}-1)(Y_i-{\nu_i})
\end{align}  

 Let $W_{ij}$ denote a $j$th component of a vector $\bW_i$. Then, \begin{align}
\label{eqn:10}
\frac{1}{\sqrt{n}}\sum_{i=1}^{n}W_{ij}(\frac{C_i}{\hat{\delta_i}}-1)(Y_i-\hat{\nu_i})&
=\frac{1}{\sqrt{n}}\sum_{i=1}^{n}W_{ij}(\frac{C_i}{\hat{\delta_i}}-\frac{C_i}{\delta_i})(\nu_i-\hat{\nu_i})\\ \nonumber
&+\frac{1}{\sqrt{n}}\sum_{i=1}^{n}W_{ij}(\frac{C_i}{{\delta_i}}-1)(\nu_i-\hat{\nu}_i)+\frac{1}{\sqrt{n}}\sum_{i=1}^{n}W_{ij}C_i(\frac{1}{\hat{\delta_i}}-\frac{1}{{\delta_i}})(Y_i-\nu_i) \\ \nonumber
&+\frac{1}{\sqrt{n}}\sum_{i=1}^{n}W_{ij}(\frac{C_i}{{\delta_i}}-1)(Y_i-{\nu_i})
\end{align}  

The limit properties of each component in equation (\ref{eqn:10}) are as follows.

For a first term, by Assumption \ref{assumption} and a Cauchy-Schwarz inequality,  we have
\begin{align}
\label{eqn:10.1}
\frac{1}{\sqrt{n}}\sum_{i=1}^{n}W_{ij}(\frac{C_i}{\hat{\delta_i}}-\frac{C_i}{\delta_i})(\nu_i-\hat{\nu_i})&\nonumber=\frac{1}{\sqrt{n}}\sum_{i=1}^{n}W_{ij}\frac{C_i}{\delta_i}(\frac{\delta_i}{\hat{\delta_i}}-1)(\nu_i-\hat{\nu_i})\\ \nonumber
&\le \frac{1}{\sqrt{n}} \sqrt{\sum_{i=1}^{n} W_{ij}^2} \sqrt{\sum_{i=1}^{n} \frac{C_i^2}{\delta_i^2}(\frac{\delta_i}{\hat{\delta_i}}-1)^2(\nu_i-\hat{\nu_i})^2} \\\nonumber
& = \sqrt{\frac{1}{n}\sum_{i=1}^{n}W_{ij}^2} \sqrt{n \cdot \frac{1}{n} \sum_{i=1}^{n} \frac{C_i^2}{\delta_i^2}(\frac{1}{\hat{\delta_i}}-\frac{1}{\delta})^2(\nu_i-\hat{\nu_i})^2} \\\nonumber
& \le \sqrt{\frac{1}{n}\sum_{i=1}^{n} W_{ij}^2} \sqrt{n \cdot \|\frac{C_i}{\delta_i}\|^2_{\infty} \cdot \|(\frac{1}{\hat{\delta_i}}-\frac{1}{\delta})(\nu_i-\hat{\nu_i})\|_2^2}
\\&=\op(1)
\end{align} for $j=1,\ldots, q$. We used  $\EE[\|\bW_i\|_2^2]=\op(1)$, $\|\frac{C_i}{\delta_i}\|_{\infty}=\op(1)$, and $\|(\frac{1}{\hat{\delta_i}}-\frac{1}{\delta})(\nu_i-\hat{\nu_i})\|_2=\op(n^{-1/2}$)

For a second term of Equation \ref{eqn:10}, since $W_{ij} (\frac{C_i}{\delta_i}-1)(\nu_i-\hat{\nu}_i)$ are i.i.d, we use CLT. \begin{align*}
\EE[W_{ij}(\frac{C_i}{{\delta_i}}-1)(\nu_i-\hat{\nu}_i)]
&=\EE[W_{ij}(\nu_i-\hat{\nu}_i)\EE[\frac{C_i}{{\delta_i}}-1\mid \bW_i, \bU_i]]
\\&=0
\end{align*} Therefore, we have
\begin{align}
\label{eqn:10.2}
\frac{1}{\sqrt{n}}\sum_{i=1}^{n}W_{ij}(\frac{C_i}{{\delta_i}}-1)(\nu_i-\hat{\nu}_i)& \nonumber =\op(\|W_{ij}(\frac{C_i}{\delta_i}-1)(\nu_i-\hat{\nu_i})\|_2) \\ \nonumber
&\le \op(\|W_{ij}\|_{\infty} \|\frac{C_i}{\delta_i}-1\|_{\infty}\|\nu_i-\hat{\nu_i}\|_2)\\&=\op(1) 
\end{align} 

Lastly, for the third term of the Equation \ref{eqn:10},  
\begin{align*}
\EE[W_{ij}C_i(\frac{1}{\hat{\delta_i}}-\frac{1}{{\delta_i}})(Y_i-\nu_i)]&=\EE[W_{ij}(\frac{1}{\hat{\delta_i}}-\frac{1}{{\delta_i}})\EE[Y_i-\nu_i\mid \bW_i, \bU_i] \EE[C_i\mid \bW_i, \bU_i]]
\\&=0\end{align*} by Assumption \ref{assumption} (a). Therefore, \begin{align}
\label{eqn:10.3}
\frac{1}{\sqrt{n}}\sum_{i=1}^{n}W_{ij}C_i(\frac{1}{\hat{\delta_i}}-\frac{1}{{\delta_i}})(Y_i-\nu_i)&=\op(\| W_{ij}\frac{C_i}{\delta_i} (\frac{\delta_i}{\hat{\delta_i}}-1)(Y_i-\nu_i) \|_{2}) \nonumber
\\ &\le \op(\|W_{ij}\|_{\infty}\|\frac{\delta_i}{\hat{\delta_i}}-1\|_2 \|Y_i-\nu_i\|_{\infty})=\op(1).
\end{align}
Putting Eqations \ref{eqn:10.1}, \ref{eqn:10.2}, \ref{eqn:10.3} together, we have
 \begin{align*}
\frac{1}{\sqrt{n}}\sum_{i=1}^{n}W_{ij}(\epsilon_i+(\frac{C_i}{\hat{\delta_i}}-1)(Y_i-\hat{\nu_i}))=\frac{1}{\sqrt{n}}\sum_{i=1}^{n}(W_{ij}(\epsilon_i+(\frac{ C_i}{{\delta_i}}-1)(Y_i-{\nu_i}))+\op(1).
\end{align*}  which naturally leads to
 \begin{align}
 \label{eqn:11}
\frac{1}{\sqrt{n}}\sum_{i=1}^{n}\bW_{i}(\epsilon_i+(\frac{C_i}{\hat{\delta_i}}-1)(Y_i-\hat{\nu_i}))=\frac{1}{\sqrt{n}}\sum_{i=1}^{n}(\bW_{i}(\epsilon_i+(\frac{ C_i}{{\delta_i}}-1)(Y_i-{\nu_i}))+\op(1).
\end{align}  
The remaining task is to get a limiting distribution of Equation \ref{eqn:11}.
Since $\bW_{i}(\epsilon_i+(\frac{C_i}{{\delta_i}}-1)(Y_i-{\nu_i}))$ are i.i.d variable with mean zero, its variance is 
 \begin{align}
\EE[&(\epsilon_i^2+(\frac{C_i}{\delta_i}-1)^2(Y_i-{\nu_i})^2+2\epsilon_i(\frac{C_i}{\delta_i}-1)(Y_i-{\nu_i}))\bW_i^T\bW_i] \nonumber \\ 
=&\EE[\epsilon_i^2\bW_i^T\bW_i]+\EE[\EE[(\frac{C_i}{\delta_i}-1)^2(Y_i-{\nu_i})^2+2\epsilon_i(\frac{C_i}{\delta_i}-1)(Y_i-{\nu_i}))\mid \bW_i, \bU_i]\bW_i^T\bW_i] \nonumber \\
=&\EE[\epsilon_i^2\bW_i^T\bW_i]+\EE[\EE[(\frac{C_i}{\delta_i}-1)^2\mid \bW_i, \bU_i]\EE[(Y_i-{\nu_i})^2\mid \bW_i, \bU_i]\bW_i^T\bW_i] \nonumber\\
&+2\EE[\EE[\epsilon_i(Y_i-{\nu_i})\mid \bW_i, \bU_i]\EE[\frac{C_i}{\delta_i}-1\mid \bW_i, \bU_i]\bW_i^T\bW_i] \nonumber \\
=&\EE[\epsilon_i^2\bW_i^T\bW_i]+\EE[(\frac{1}{\delta_i}-1)(Y_i-{\nu_i})^2\bW_i^T\bW_i]
\label{eqn:12}
\end{align}
Therefore, by CLT, 
\begin{equation} \frac{1}{\sqrt{n}}\sum_{i=1}^{n}\bW_i\left(\epsilon_i+(\frac{C_i}{\delta_i}-1)(Y_i-\hat{\nu_i})\right)\xrightarrow{D} \cN(0, \EE[(\epsilon_i^2+(\frac{1}{\delta}-1)(Y_i-{\nu_i})^2)\bW_i^T\bW_i] )\label{part2}
\end{equation} holds. Then we plug in the result of Equation $\ref{part1}, \ref{part2}$ to Equation $\ref{part0}$ and apply Slutsky Theorem to get the desired result.\\

\noindent {\bf Part 2: Variance consistency}


Let us denote $\epsilon^*_i=\epsilon_i+(\frac{C_i}{\delta_i}-1)(Y_i-v_i)$ and $\hat{\epsilon}^*_i=\hat{Y}_i^{UW}-\bW_i^T\hat{\bbeta}$. We claim that $\|\Sigma-\hat{\Sigma}\|_2=\op(1)$, where
\begin{align*}
\hat{\Sigma}&=(\frac{1}{n}\sum_{i=1}^{n}\bW_i\bW_i^T)^{-1}\frac{1}{n}\sum_{i=1}^{n}(\hat{\epsilon}_i^{*2} \bW_i\bW_i^T)(\frac{1}{n}\sum_{i=1}^{n}\bW_i\bW_i^T)^{-1}\\
\Sigma&=\EE[\bW_i\bW_i^T]^{-1}\EE[\epsilon_i^{*2}\bW_i\bW_i^T]\EE[\bW_i\bW_i]^{-1}
\end{align*}
Note that the above expression for $\Sigma$ use the fact that $\EE[\epsilon_i^{*2}\bW_i\bW_i^T]=\EE[(\epsilon_i^2+(1-\frac{1}{\delta_i})(Y_i-v_i)^2)\bW_i\bW_i^T]$ whose derivation is in Equation \ref{eqn:12}.

Then,\begin{align}
\label{eqn:20}
\hat{\Sigma}_{UW}-\Sigma_{UW}&=(\frac{1}{n}\sum_{i=1}^{n}\bW_i\bW_i^T)^{-1}\{ \frac{1}{n}\sum_{i=1}^{n}({\hat{\epsilon}_i^{*2}} \bW_i\bW_i^T)-\EE[\epsilon_i^{*2}\bW_i\bW_i^T]\} (\frac{1}{n}\sum_{i=1}^{n}\bW_i\bW_i^T)^{-1} \nonumber\\&-\{ \EE[\bW_i\bW_i^T]^{-1}-(\frac{1}{n}\sum_{i=1}^{n}\bW_i\bW_i^T)^{-1}\} \EE[\epsilon_i^{*2}\bW_i\bW_i^T]\{ \EE[\bW_i\bW_i^T]^{-1}-(\frac{1}{n}\sum_{i=1}^{n}\bW_i\bW_i^T)^{-1}\}
\end{align}

 First we show that $\EE[\epsilon^{*2}_iW_{ij}W_{ik}]$ is bounded for $j, k \in \{1,...,q\}$; 
 \begin{align}
 \label{eqn:20.1}
\EE[\epsilon^{*2}_iW_{ij}W_{ik}]&=\EE[(\epsilon_i+(\frac{C_i}{\delta_i}-1)(Y_i-v_i))^2W_{ij}W_{ik}]\nonumber\\
&=\EE\left[(\epsilon_i^2+(\frac{C_i}{\delta_i}-1)^2(Y_i-v_i)^2+2\epsilon_i(\frac{C_i}{\delta_i}-1)(Y_i-v_i)W_{ij}W_{ik}\right] \nonumber \\
&=\EE[\epsilon_i^2W_{ij}W_{ik}]+\EE[(\frac{1}{\delta_i}-1) \EE[(Y_i-v_i)^2\mid \bW_i, \bU_i]W_{ij}W_{ik}]
\end{align} Each term in expectation is bounded by Assumption \ref{assumption}. Combining this to the fact that  $\EE[\bW_i\bW_i^T]^{-1}-(\frac{1}{n}\sum_{i=1}^{n}\bW_i\bW_i^T)^{-1}=\op(1)$, second term of the Equation \ref{eqn:20} is $\op(1)$. Also, $(\frac{1}{n}\sum_{i=1}^{n}\bW_i\bW_i^T)^{-1}$ is bounded by a full-rank assumption of $\EE[\bW_i \bW_i^T]$. Therefore, it is suffice to show that $\frac{1}{n}\sum_{i=1}^{n}(({\hat{\epsilon}_i^{*2}} \bW_i\bW_i^T)-\EE[\epsilon_i^{*2}\bW_i\bW_i^T])=\op(1)$. 

To this end, we show that every component of them are $\op(1)$. That is, for $j,k=1,..,q$, we claim that 
\begin{align}\label{eqn:21}
    \left|\frac{1}{n}\sum_{i=1}^{n}({\hat{\epsilon}_i^{*2}} W_{ij}W_{ik})-\EE[\epsilon_i^{*2}W_{ij}W_{ik}]\right|=\op(1).
\end{align} 
A left-hand side of Equation \ref{eqn:21} can be bounded by 
\begin{align}
\label{eqn:22}|\frac{1}{n}\sum_{i=1}^{n}&({\hat{\epsilon}_i^{*2}} W_{ij}W_{ik})-\EE[\epsilon_i^{*2}W_{ij}W_{ik}]| \nonumber \\& \le |\frac{1}{n}\sum_{i=1}^{n}({\hat{\epsilon}_i^{*2}} W_{ij}W_{ik}-\epsilon_i^{*2}W_{ij}W_{ik})|+|\frac{1}{n}\sum_{i=1}^{n}(\epsilon_i^{*2}W_{ij}W_{ik})-\EE[\epsilon_i^{*2}W_{ij}W_{ik}]|
\end{align}

Second term of Equation \ref{eqn:22} is $\op(1)$ by a law of large number. Therefore, it is suffice to show that the first term is $\op(1)$. 

The first term can be bounded by
\begin{align}
|\frac{1}{n}\sum_{i=1}^{n}({\hat{\epsilon}_i^{*2}} &W_{ij}W_{ik}-\epsilon_i^{*2}W_{ij}W_{ik})| \nonumber \\
&=|\frac{1}{n}\sum_{i=1}^{n}({\hat{\epsilon}_i^*-{\epsilon}_i^*})({\hat{\epsilon}_i^*+{\epsilon}_i^*}) W_{ij}W_{ik})| \nonumber \\
&= |\frac{1}{n}\sum_{i=1}^{n}\{ ({\hat{\epsilon}_i^*-{\epsilon}_i^*})W_{ij}\} \{ ({\hat{\epsilon}_i^*-{\epsilon}_i^*+2{\epsilon}_i^*})W_{ik}  \} |\nonumber \\
&\le \sqrt{\frac{1}{n}\sum_{i=1}^{n}\{ ({\hat{\epsilon}_i^*-{\epsilon}_i^*})W_{ij}\}^2 } \sqrt{ \frac{1}{n}\sum_{i=1}^{n}\{ ({\hat{\epsilon}_i^*-{\epsilon}_i^*+2{\epsilon}_i^*})W_{ik}  \}^2 } 
\label{eqn:23}
\end{align} where the last step uses a Cauchy-Shwarz inequality. 

Using a decomposition
 \begin{align}
\label{eqn:24}
\hat{\epsilon}_i^*-\epsilon_i^*&=\hat{Y}_i^{UW}-\bW_i^T\hat{\bbeta}-\epsilon_i^* \nonumber \\
&=Y_i+(\frac{C_i}{\hat{\delta_i}}-1)(Y_i-\hat{\nu}_i)-\bW_i^T\hat{\bbeta}-\epsilon_i-(\frac{C_i}{\delta_i}-1)(Y_i-v_i) \nonumber\\
&=\bW_i^T(\bbeta-\hat{\bbeta})+(\frac{C_i}{\hat{\delta_i}}-1)(Y_i-\hat{\nu}_i)-(\frac{C_i}{\delta_i}-1)(Y_i-v_i) \nonumber\\
&=\bW_i^T(\bbeta-\hat{\bbeta})+\left(1-\frac{\delta_i}{\hat{\delta_i}}\right)(\hat{\nu}_i-{\nu}_i)\nonumber,
\end{align} 
and a Cauchy-Shwarz inequality, we have
\begin{align}
\frac{1}{n}\sum_{i=1}^{n} ({\hat{\epsilon}_i^*-{\epsilon}_i^*})^2 & \le  \frac{1}{n}\sum_{i=1}^{n} (\bW_i^T(\bbeta-\hat{\bbeta})+ \left(1-\frac{\delta_i}{\hat{\delta_i}}\right)(\hat{\nu}_i-{\nu}_i))^2 \nonumber \\ 
& \le  \| \bbeta-\hat{\bbeta}\|_2^2 \frac{1}{n}\sum_{i=1}^{n} \|\bW_i^T\|_2^2 + \frac{1}{n}\sum_{i=1}^{n}\left(1-\frac{\delta_i}{\hat{\delta_i}}\right)^2(\hat{\nu}_i-{\nu}_i)^2 \nonumber \\
&+2\| \bbeta-\hat{\bbeta}\|_2 \frac{1}{n}\sum_{i=1}^{n} \|\bW_i^T\|_2\left(1-\frac{\delta_i}{\hat{\delta_i}}\right)(\hat{\nu}_i-{\nu}_i) \nonumber \\
&=\op(1)\,,
\end{align}
where the last step uses the fact that
$\|\bbeta-\hat\bbeta\|_2=\op(1)$, $\frac{1}{n}\sum_{i=1}^n\|\bW_i\|_2^2\le p\|\bW_i\|_\infty^2$, which is bounded by a constant, and that $(1-\delta_i/\hat\delta_i)^2(\hat \nu_i-\nu_i)^2$ has vanishing $L_1$ norm for each $i$ according to Assumption \ref{assumption}(e).

Then, since $\|\bW_i\|_{\infty}$ bounded, \begin{equation} 
\label{eqn:25} 
\sqrt{\frac{1}{n}\sum_{i=1}^{n}\{ ({\hat{\epsilon}_i^*-{\epsilon}_i^*})W_{ij}\}^2 }=\op(1). 
\end{equation} 

Also, applying $(\hat{\epsilon}_i^*-\epsilon_i^*)^2=\op(1)$, we have \begin{align}
 \sqrt{ \frac{1}{n}\sum_{i=1}^{n}\{ ({\hat{\epsilon}_i^*+{\epsilon}_i^*})W_{ik}  \}^2 }& \le \sqrt{ \frac{1}{n}\sum_{i=1}^{n} \{2(\hat{\epsilon}_i^*-\epsilon_i^*)^2W_{ik}^2+2(\epsilon_i^*)^2W_{ik}^2\}} \nonumber \\
 &=\Op(1)
 \label{eqn:26}
\end{align}
because $\frac{1}{n}\sum_{i=1}^n (\hat{\epsilon_i^*}-\epsilon_i^*)^2W_{ik}^2=\op(1)$ as Equation \ref{eqn:25} and $\frac{1}{n}\sum_{i=1}^n (\epsilon_i^*)^2W_{ik}^2=\Op(1)$ by Assumption \ref{assumption}(d).
Plugging in the Equation \ref{eqn:25}, \ref{eqn:26} to Equation \ref{eqn:23} completes the proof.

\end{proof}

\subsection{Proof of \Cref{thm:high_better} }

\begin{proof}
\begin{align*}
\Sigma_{U}-\Sigma_{UW}&=\EE[\bW_i\bW^T_i]^{-1}\EE[(\epsilon_i^2+(\frac{1}{\delta_i}-1)(Y_i-\mu_i)^2)\bW_i\bW_i^T]\EE[\bW_i\bW^T_i]^{-1}\\
&-\EE[\bW_i\bW^T_i]^{-1}\EE[(\epsilon_i^2+(\frac{1}{\delta_i}-1)(Y_i-\nu_i)^2)\bW_i\bW_i^T]\EE[\bW_i\bW^T_i]^{-1}\\
&=\EE[\bW_i\bW^T_i]^{-1}\EE[(\frac{1}{\delta_i}-1)((Y_i-\mu_i)^2-(Y_i-\nu_i)^2)\bW_i\bW_i^T]\EE[\bW_i\bW^T_i]^{-1}
\end{align*}

Since
\begin{align*}
\EE[(Y_i-\nu_i)(\nu_i-\mu_i)|\bW_i]&=\EE_{\bU_i}[\EE[(Y_i-\nu_i)(\nu_i-\mu_i)|\bW_i,\bU_i]]\\
&=\EE_{\bU_i}[(\nu_i-\mu_i)\EE[(Y_i-\nu_i)|\bW_i,\bU_i]]\\
&=0,
\end{align*} by Assumption \ref{assumption}(c), 
we have 

\begin{align*}
\EE[(Y_i-\mu_i)^2-(Y_i-\nu_i)^2|\bW_i]&=\EE[\{(Y_i-\nu_i)+(\nu_i-\mu_i)\}^2-(Y_i-\nu_i)^2|\bW_i]\\
&=\EE[(\nu_i-\mu_i)^2|\bW_i]
\end{align*}

Combining with $\EE[\bW_i\bW^T_i] \succcurlyeq 0$ and $(\frac{1}{\delta_i}-1)(\nu_i-\mu_i)^2\ge 0$, we have 

\begin{align*}
\Sigma_{U}-\Sigma_{UW}&=\EE[\bW_i\bW^T_i]^{-1}\EE[(\frac{1}{\delta_i}-1)(\nu_i-\mu_i)^2\bW_i\bW_i^T]\EE[\bW_i\bW^T_i]^{-1}
\\& \succcurlyeq 0
\end{align*}

\end{proof}

\clearpage
\section{Imputation methods}\label{sec:impute-method}

    \subsection{Probabilistic modeling of peptide datasets} 
    The semiparametric inference results established in the main paper allow us to use more flexible non-parametric machine learning and deep learning models to estimate the mean regression nuisance function and improve the imputation quality. 
    Inspired by recent advancements in conditional variational inference \citep{vae,sohn2015learning,ivanov2018variational} in the machine learning community, \citet{Du:2022} propose a variational autoencoder (VAE) model for imputation of single-cell multi-omics data by utilizing a masking procedure to inform the missing patterns and help the model to learn conditional distributions among features, which we refer to VAEIT in the current section. 
    
    Specifically, VAEIT models the missing features as a conditional probability estimation problem. 
    For each individual, we denote its measurements of $p$ peptides by a random vector $\bY=(Y_1,\ldots,Y_p)\in\RR^p$. We introduce a binary mask $\bM\in \{0,1\}^p$ for $\bY$ and its bitwise complement $\bM^c$, such that the $j$th entry of the observed sample $\bY_{M^c}$ is $Y_j$ if $M_j=1$ and $0$ otherwise.
    We use an authentic missing pattern $\bM_a = \mathbf{1}_{p} - \bC$ to represent which components of $\bY$ are actually missing, while the distribution of $\bM$ can be arbitrary during training. 
    For example, if we want to model missing completely at random, the entries of $\bM$ could be independent Bernoulli random variables.
    Furthermore, we can incorporate extra structural information to model the situation of missing modality.
    To model the conditional distribution of the missing peptides given the observed values, we consider the following maximum likelihood problem:
        \begin{align*}
            \max\limits_{\theta}\EE_{\bY,\bM}\log p_{\theta}(\bY_{\bM}\mid \bY_{\bM^c}, \bM, \bW).
        \end{align*}
        In other words, we aim to determine the conditional distribution of $\bY_\bM$ given $\bY_{\bM^c}$, $\bM$ and the low-dimensional covariate $\bW\in\RR^q$.
        We utilize the flexibility of neural networks to jointly model all conditional distributions at once. 
        
        Because the above condition density itself is hard to formulate and optimize, we follow the variational Bayesian approach \citep{blei2017variational}
        to maximize the negative evidence lower bound (ELBO):
        \begin{align}
            \log p_{\theta}(\bY_\bM\mid \bY_{\bM^c}, \bM, \bW) &\geq\underbrace{\EE_{q_{\psi}(\bZ \mid \bY,\bM, \bW)}\log p_{\theta_2}(\bY_\bM\mid \bZ,\bY_{\bM^c}, M, \bW)}_{\mathcal{L}_{\text{impute}}} \notag\\
            &\qquad - \beta_{\text{kl}}\cdot KL(q_{\psi}(\bZ\mid \bY,\bM, \bW)\| p_{\theta_1}(\bZ\mid \bY_{\bM^c}, \bM, \bW) ) =: \cL_{M}, \label{eq:elbo}
        \end{align}
        where $\bZ\in\RR^m$ is a latent variable with approximate posterior distribution $q_{\psi}$, $\beta_{\text{kl}}=1$ is the regularization strength, $KL$ denotes the Kullback–Leibler divergence, and $\theta=(\theta_1,\theta_2)$.
        Increasing the regularization strength $\beta_{\text{kl}}$ usually improves the representation learning, which gives rise to the so-called $\beta$-VAE. 
        We specify the distributions for data as follows.
        
        Under the target distribution $p_{\theta_1}$, we assume that the latent variables are normally distributed:
        \begin{align}
            \bZ\mid \bY_{M^c},\bM, \bW \sim \cN(\mu_{\theta_1}, \diag(\sigma_{\theta_1,1}^2,\ldots,\sigma_{\theta_1,m}^2)).
        \end{align}
        Ideally, we want $\bZ$ generated from $p_{\theta_1}$ to be as close as possible to the one generated from the proposal distribution $q_{\psi}$ when $\bY$ is fully observed except for its authentic missing entries $\bM_a = \mathbf{1}_{p}-\bC$:
        \begin{align}
        \bZ\mid \bY_{\bM_a^c},\bM_a, \bW \sim \mathcal{N}(\mu_{\psi}, \mathrm{diag}(\sigma_{\psi,1}^2,\ldots,\sigma_{\psi,m}^2)).
        \end{align}
        This formulation also allows us to compute the KL divergence analytically in the ELBO \eqref{eq:elbo}, while it is possible to extend to normal mixtures to model more complex latent structures \citep{du2020model}.
        In our implementation, we simply set $q_{\psi}(\bZ\mid \bY_{\bM_a^c},\bM_a, \bW)=p_{\theta_1}(\bZ\mid \bY_{\bM^c},\bM, \bW)$ to reduce computational complexity.
        Finally, $q_{\psi}$ and $p_{\theta_2}$ are modeled as two fully-factorized Gaussian distributions, whose mean and variance are estimated by two neural networks, respectively.
        The generative distribution $p_{\theta_1}$ are also assumed to be fully-factorized for $\bY_\bM$ given $\bZ,\bY_{\bM^c}$, $\bM$ and $\bW$.
        We use normal distributions to model the peptide abundances.
        We assume that the intensities are generated based on $\bZ$ as follows
        \begin{align}
            Y_{j} \mid \bZ,\bM, \bW &\sim \cN(\lambda_{j},\theta_j),
        \end{align}
        which are independent of $\bM$ given $\bZ$.
        Here the parameters $\lambda_{j}$ and $\theta_j$ are the expected intensity and the variance of the normal distribution.
    	The posterior expectations $\lambda_{j}$'s are outputted by the decoder and sample-specific, while the dispersion parameters $\theta_j$'s are treated as trainable variables.
        These parameters are learned from the data.
        
        The aforementioned probabilistic modeling \eqref{eq:elbo} emphasizes missing features imputation. On the other hand, we not only want to impute the unobserved quantities but also denoise the observed quantities.
        Therefore, we also attempt to maximize the reconstruction likelihood
        \begin{align}
    	    \cL_{rec} &:= \EE_{p_{\theta_2}(\bZ\mid \bY_{\bM^c}, \bM)}\log p_{\theta_1}(\bY_{\bM^c}\mid \bZ,\bM, \bW). \label{eq:obj2}
        \end{align}

    \subsection{Network architecture}
        VAEIT is implemented using the Tensorflow (version 2.4.1) Python library \citep{tensorflow2015-whitepaper}.
        VAEIT consists of three main branches, the mask encoder, the main encoder, and the main decoder.
        For each sample, a missing mask $\bM$ is embedded as $\mathbf{E}$ to a dense vector of dimension 128 through the mask encoder, which greatly reduces the input dimension to the main encoder and decoder.
        Then, the encoder takes data $\bY$ (log-normalized peptide abundance), a mask embedding vector $\mathbf{E}$, and (optional) covariates $\bW$ as input, and outputs the estimated posterior mean and variance of the distribution of the latent variable $\bZ$.
        Next, a realization is drawn from this posterior distribution and fed to the decoder along with the mask embedding vector $\mathbf{E}$ and the low-dimensional covariates $\bW$.
        The decoder finally outputs the posterior mean of $\bY$.

        The encoder has two hidden layers of 64 and 16 units, and the decoder has two hidden layers of 16 and 64 units. 
        The activation functions are set to LeakyReLU with parameter 0.2.
        The latent dimension is set to be 4.

        \subsection{Model training}
    VAEIT is trained in an end-to-end manner.
            The objective function is a convex combination of the ELBO \eqref{eq:elbo} and the reconstruction likelihood \eqref{eq:obj2}:
            $$\cL:=\beta_{\text{unobs}} \cL_{M}+(1-\beta_{\text{unobs}})\cL_{recon} ,$$
            where $\beta_{\text{unobs}}\in[0,1]$ is a hyperparameter set to be 0.9 for all experiments.
            We set the KL regularization parameter as $\beta_{\text{kl}} = 10$.
            The parameters are optimized by Monte Carlo sampling to maximize the weighted average of the reconstruction likelihood and the imputation likelihood while minimizing the KL divergence between masked posterior latent variable $\bZ\mid \bY_{\bM^c},\bM, \bW$ and the authentic posterior latent variable $\bZ\mid \bY_{\bM_a^c},\bM_a, \bW$.
            During training, with equal probability, we observe the original data and the masked data.
            The mask is repeatedly randomly generated for each sample at the beginning of every gradient update step in each epoch during the optimization process, such that each modality is observed with equal probability, and each entry is further randomly masked out with probability 0.5.            
            The default variable initializer in Tensorflow is used, sampling the weight matrix from a uniform distribution and setting bias vectors to be zero.
            We trained our model for 300 epochs using the AdamW optimizer \citep{loshchilov2017decoupled} with full batches and a learning rate of 1e-3 and a weight decay of 1e-4.
            We also use batch normalization to aid in training stability.

\clearpage
\pagebreak

\section{Supplementary figures and tables}
\textcolor{white}{.}

\begin{figure}[H]
\centering
	\includegraphics[width=0.6\textwidth]{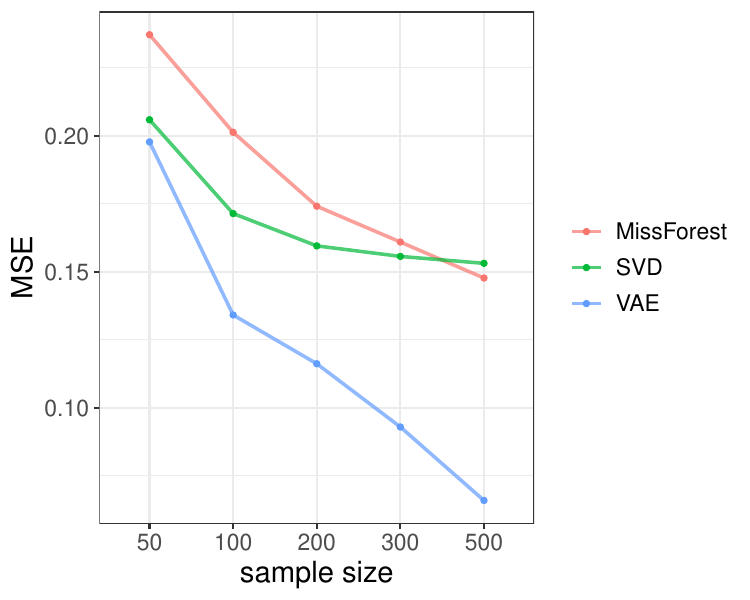}
\caption{ Mean squared error of the three imputation methods in the setting of Model 3. The numbers are averaged over 200 repetitions (Section 3.2).} 
\label{fig:sim3_mse }
\end{figure}

\begin{figure}[H]
\centering
	\includegraphics[width=13cm]{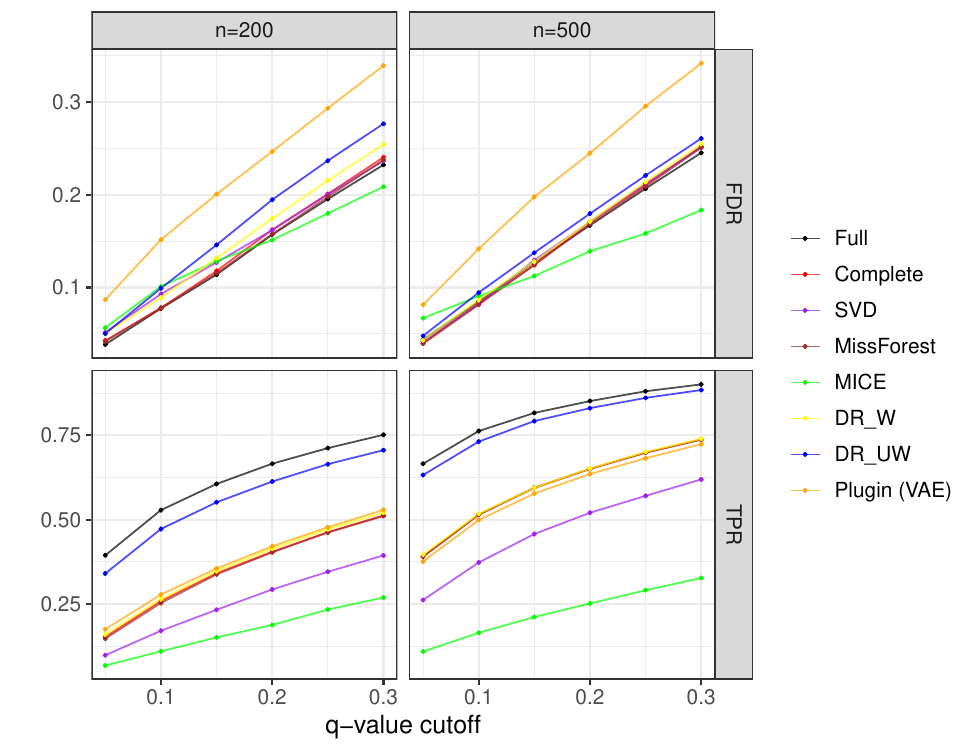}
\caption{Simulation result of Model 1 (Section 3.2). } 
\label{fig:sim1}
\end{figure}

\clearpage

\begin{figure}[H]
\centering
	\includegraphics[width=14cm]{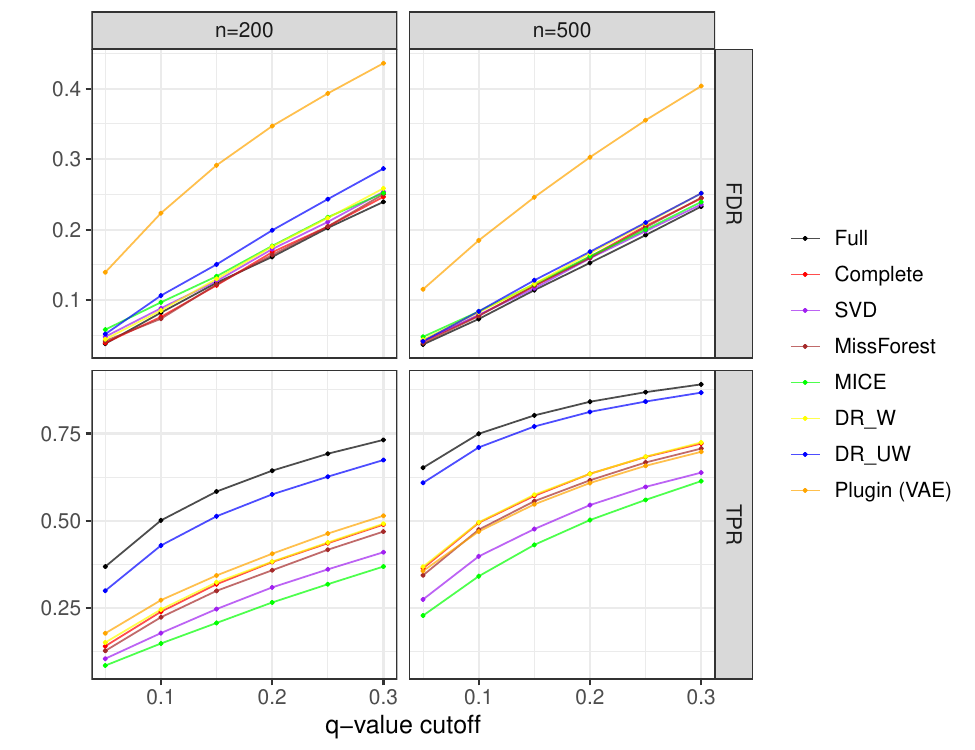}
\caption{Simulation result of Model 2 (Section 3.2). } 
\label{fig:sim2}
\end{figure}

\begin{figure}[H]
\centering
	\includegraphics[width=14cm]{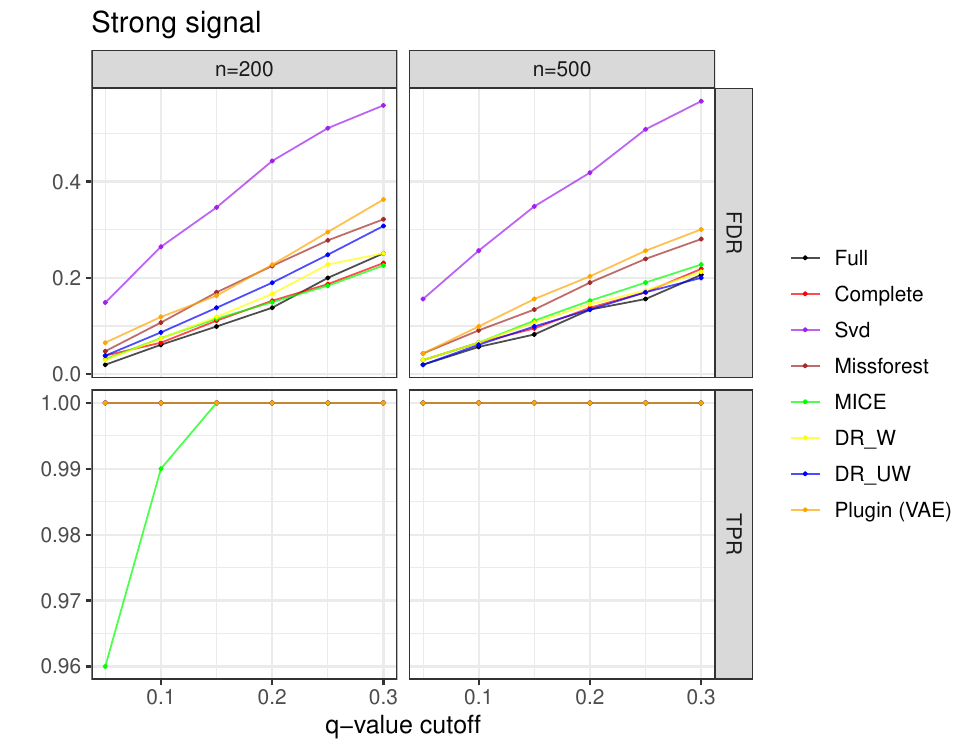}
\caption{Simulation result of Model 3 under strong signals (Section 3.2). } 
\label{fig:sim3_strong}
\end{figure}

\begin{figure}[H]
\centering
	\includegraphics[width=14cm]{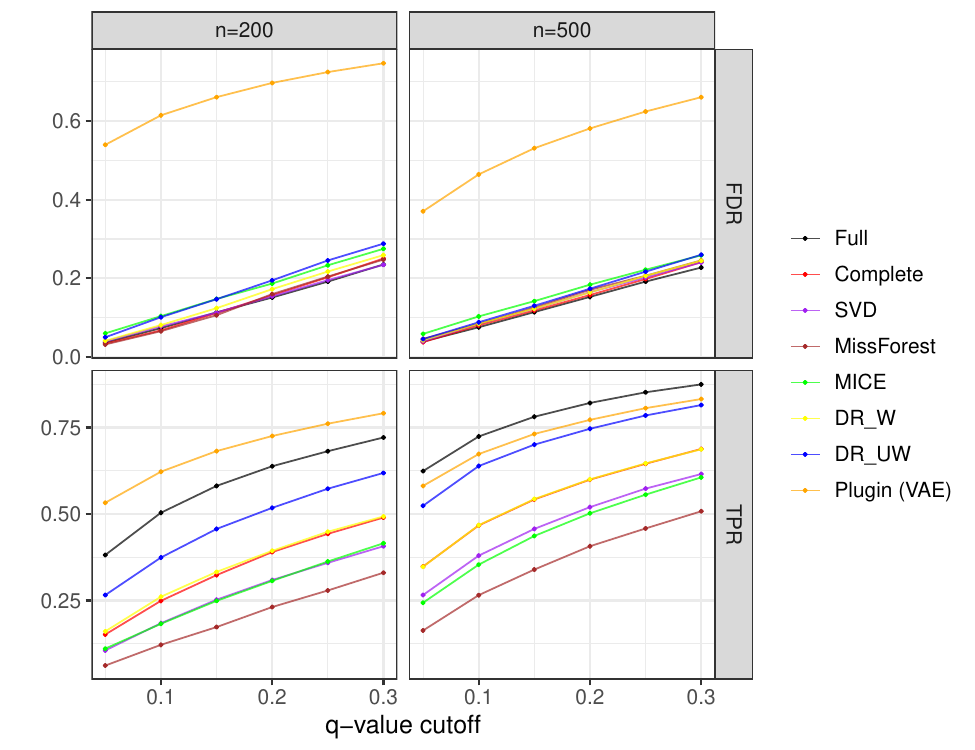}
\caption{Simulation result of Model 4 (Section 3.2). } 
\label{fig:sim4}
\end{figure}

\clearpage

\begin{figure}[h]
\centering
	\includegraphics[width=10cm]{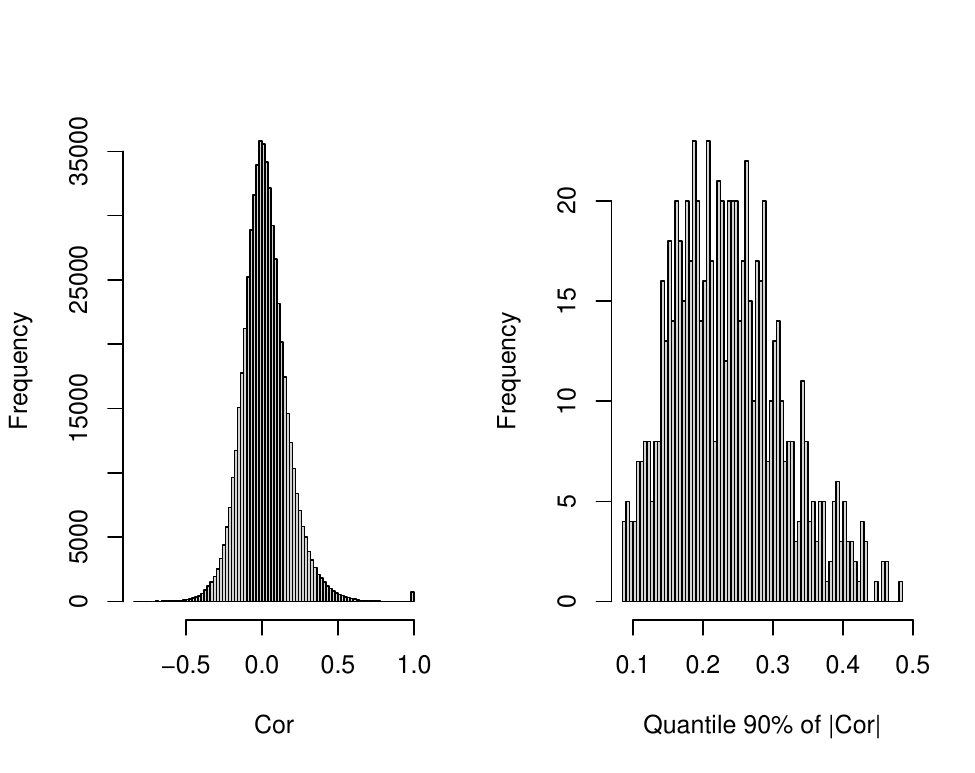}
\caption{Histogram of correlation coefficient between peptides (left) and a 90\% quantile absolute value of correlation coefficients computed for each peptide (right) (Section 4)} 
\label{fig:correlation}
\end{figure}

\begin{figure}[h]
\centering
	\includegraphics[width=11cm]{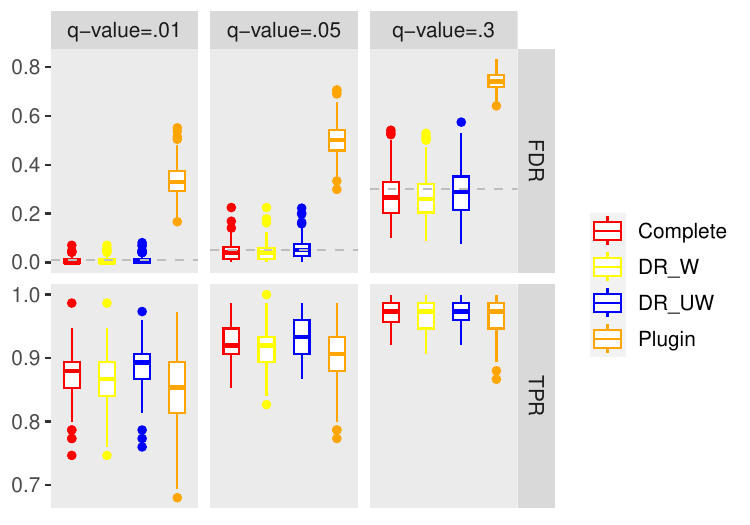}
\caption{A result of realistic simulation with a single-cell dataset (Setting 1); FDR and TPR are summarized under different q-value cutoffs (0.01, 0.05 and 0.3) (Section 4)} 
\label{fig:semi_null_signal_1}
\end{figure}

\begin{figure}[h]
\centering
	\includegraphics[width=11cm]{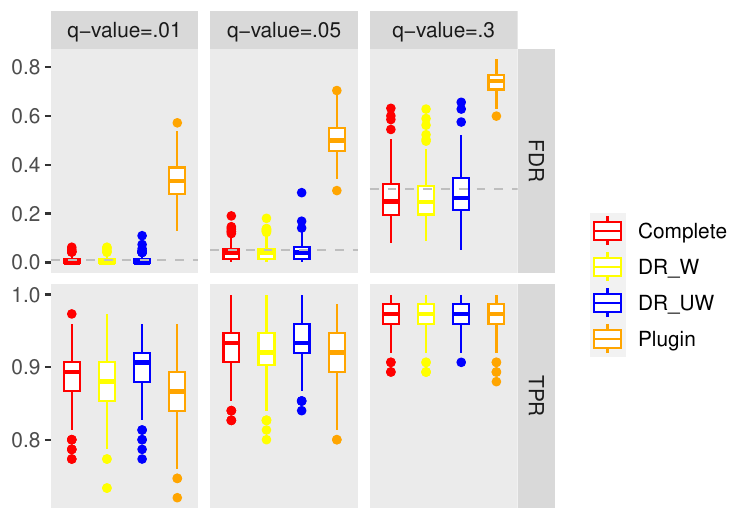}
\caption{A result of realistic simulation with a single-cell dataset (Setting 2); FDR and TPR are summarized under different q-value cutoffs (0.01, 0.05 and 0.3) (Section 4)} 
\label{fig:semi_null_signal_2}
\end{figure}

\begin{table}[h]
\centering
\begin{tabular}{|c|c|c|c|c|c|c|}
\hline
\multirow{2}{*}{q-value cutoff} & \multicolumn{3}{c|}{\% Protein overlaps}                           & \multicolumn{3}{c|}{Number of additional peptides}                 \\ \cline{2-7} 
                        & {DR\_W} & {DR\_UW} & Plugin & {DR\_W} & {DR\_UW} & Plugin \\ \hline
0.01                    & {1}    & {0.90}   & 0.51    & {6}     & {31}     & 179     \\ \hline
0.05                    & {0.90}  & {0.71}   & 0.44    & {62}     & {104}     & 257     \\ \hline
0.1                     & {0.74}     & {0.62}   & 0.41    & {100}     & {143}     & 288     \\ \hline
0.3                     & {0.51}     & {0.47}   & 0.38    & {214}     & {258}     & 344     \\ \hline
\end{tabular}
\caption{\rm The proportions of peptides, whose corresponding proteins are included in protein lists corresponding to the peptides selected by the Complete method with a q-value cutoff of 0.05. Only the peptides additionally selected by each method compared to the Complete method (with a q-value cutoff of 0.01) are considered. A threshold 0.7 is applied to the observation rate of peptides (Section 4). }
\label{table:proportions}
\end{table}

\clearpage

\begin{figure}[h]
\centering
	\includegraphics[width=13cm]{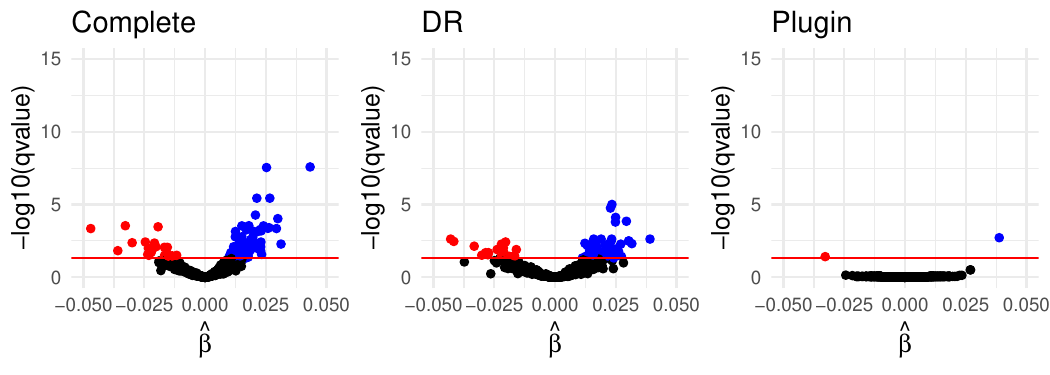}
\caption{Volcano plot with a completely noisy imputation (a q-value cutoff=0.05) (Section 6)} 
\label{fig:garbage}
\end{figure}

\begin{figure}[h]
\centering
	\includegraphics[width=7cm]{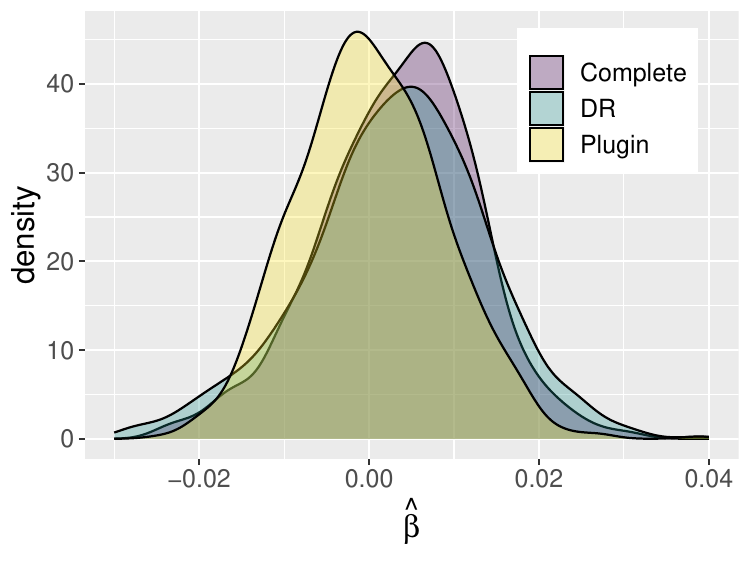}
\caption{Distribution of estimated coefficient $\hat{\beta}$ for the diameter variable under a completely noisy imputation (Section 6)} 
\label{fig:garbage_betadist}
\end{figure}

\end{document}